\newif\ifsingle

\newif\ifFullVersion
\FullVersiontrue 

\ifsingle
\documentclass[11pt,draftclsnofoot, onecolumn]{IEEEtran}		
\else		
\documentclass[10pt,final, twocolumn]{IEEEtran}
\fi

\usepackage{RTSNet}
\usepackage{STSP}

\definecolor{NewColor}{rgb}{0,0,0}

\title{RTSNet: Learning to Smooth in Partially Known State-Space Models (Preprint)
}
\author{  
	\IEEEauthorblockN{Guy Revach~\IEEEmembership{Graduate Student Member, ~IEEE}, Xiaoyong Ni~\IEEEmembership{Student Member, ~IEEE}, Nir Shlezinger~\IEEEmembership{Senior Member,~IEEE}, Ruud J. G. van Sloun~\IEEEmembership{Member,~IEEE}, and Yonina C. Eldar~\IEEEmembership{Fellow,~IEEE}\\ 
	} 
	\thanks{ 
This manuscript is a preprint. It is partially based on work presented at the IEEE International Conference on Acoustics, Speech, and Signal Processing (ICASSP) 2022~\cite{ni2021rtsnet}, and has been accepted for publication in IEEE Transactions on Signal Processing, Vol. 71, 2023.

		G. Revach and X. Ni are with the Institute for Signal and Information Processing (ISI), D-ITET, ETH Zürich, 8006 Zürich, Switzerland, 
		(e-mail: grevach@ethz.ch). 
		N. Shlezinger is with the School of ECE, Ben-Gurion University of the Negev, Beer Sheva, Israel. 
		R. J. G. van Sloun is with the EE Dpt., Eindhoven University of Technology, and with Philips Research, Eindhoven,  The Netherlands. 
		Y. C. Eldar is with the Faculty of Math and CS, Weizmann Institute of Science, Rehovot, Israel. We thank Hans-Andrea Loeliger for the helpful discussions.
	} 
	\vspace{-0.5cm}
}
\vspace{-0.75cm}

\begin{document}
	
	\maketitle
	\pagestyle{plain}
	\thispagestyle{plain}
%
%
\begin{abstract}
\textcolor{NewColor}
{The smoothing task is core to many signal-processing applications. A widely popular smoother is the \ac{rts} algorithm, which achieves minimal mean-squared error recovery with low complexity for \acl{lg} \ac{ss} models, yet is limited in systems that are only partially known, as well as \acl{nl} and non-Gaussian. 
In this work, we propose \acl{rn}, a highly efficient \acl{mb} and \acl{dd} smoothing algorithm suitable for partially known \ac{ss} models. \acl{rn} integrates dedicated trainable models into the flow of the classical \ac{rts} smoother, while iteratively refining its sequence estimate via deep unfolding methodology. As a result, \acl{rn} learns from data to reliably smooth when operating under model mismatch and nonlinearities while retaining the efficiency and interpretability of the traditional \ac{rts} smoothing algorithm. %
Our empirical study demonstrates that \acl{rn} overcomes nonlinearities and model mismatch, outperforming classic smoothers operating with both mismatched and accurate domain knowledge. Moreover, while \acl{rn} is based on compact \aclp{nn}, which leads to faster training and inference times, it demonstrates improved performance over previously proposed deep smoothers in \acl{nl} settings.}
\end{abstract}
\acresetall 
%
%
\section{Introduction}\label{sec:Intro}
A broad range of applications in signal processing and control require estimation of the hidden state of a dynamical system from noisy observations. Such tasks arise in localization, tracking, and navigation~\cite{durbin2012time}. State estimation by filtering and smoothing date back to the work of {Wiener} from 1949~\cite{wiener1949extrapolation}. Filtering (also known as \acl{rt} tracking) is the task of estimating the current state from past and current observations, while smoothing deals with simultaneous state estimation across the entire time horizon using all available data.

Arguably, the most common and celebrated filtering algorithm is the \ac{kf} proposed in the early 1960s~\cite{kalman1960new}. The \ac{kf} is a low-complexity implementation of the \ac{mmse} estimator for time-varying systems in \acl{dt} that are characterized by a linear \ac{ss} model with \ac{awgn}.  
The \ac{rts} smoother~\cite{rauch1965maximum}, also referred to here as the \ac{ks}, adapts the \ac{kf} for smoothing in \acl{dt}. 
The \ac{ks} implements \ac{mmse} estimation for \acl{lg} \ac{ss} models by applying a recursive forward pass, i.e., from the past to the future by directly applying the \ac{kf}, followed by a recursive update backward pass. 

\textcolor{NewColor}{
The \ac{ks} is given by recursive equations. These may be derived from the more general framework of message passing over factor graphs~\cite{4282128, wadehn2019state}. Alternatively, the \ac{ks} can be extended from an optimization perspective, by recasting it as a \ac{ls} system with a specific structure dictated by the \ac{ss} model, and is solved using Newton's method~\cite{humpherys2012fresh, aravkin2014optimization}. The latter perspective has further generalized \ac{eks} to \acl{nl} models. Drawing from this optimisation perspective, multiple extensions have been derived to address non-Gaussian densities (especially outliers)~\cite{aravkin2011ell, aravkin2013sparse, aravkin2014optimization}, state-dependent covariance matrices~\cite{aravkin2014smoothing}, as well as state constraints and sparsity~\cite{aravkin2014optimization, aravkin2017generalized}}. 
%
%

The \ac{ks} and its variants are \ac{mb} algorithms; \textcolor{NewColor}{that is, they assume that the underlying system's dynamics is accurately characterized by a known \ac{ss} model}. However, many \acl{rw} systems in practical use cases are complex, and it may be challenging to comprehensively and faithfully represent these systems with a fully known, tractable \ac{ss} model. Consequently, despite its low complexity and theoretical soundness, applying the \ac{ks} in practical scenarios may be limited due to its critical dependence on accurate knowledge of the underlying \ac{ss} model. \textcolor{NewColor}{Furthermore, the \acl{nl} variants of the \ac{ks}  do not share its \ac{mmse} optimality, and their performance degrades under strong nonlinearities.}
 
A common approach to deal with partially known \ac{ss} models is to impose a parametric model and then estimate its parameters. This can be achieved by {jointly} learning the parameters and state sequence using \acl{em}~\cite{ ghahramani1996parameter, dauwels2009expectation} and Bayesian probabilistic algorithms~\cite{yuen2016online,mu2017stable}, or by selecting from a set of \textit{a priori} known models~\cite{martino2017cooperative}. When training data is available, it is commonly used to tune the missing parameters in advance~\cite{xu2021ekfnet,barratt2020fitting}. These strategies are restricted to an imposed parametric model on the underlying dynamics (e.g., linear models with Gaussian noises), and thus may still lead to mismatched operation. Alternatively, uncertainty can be managed through the use of Bayesian probabilistic algorithms as in~\cite{yuen2016online,mu2017stable}, by selecting from a set of a priori known models as in~\cite{martino2017cooperative}, or by implementing robust estimation methods as in~\cite{zorzi2017robustness,longhini2021learning}. These techniques are typically designed for the worst-case deviation between the postulated model and the ground truth, rarely approaching the performance achievable with full domain knowledge.
%
%

\Ac{dd} approaches are an alternative to \ac{mb} algorithms, relaxing the requirement for explicit and accurate knowledge of the underlying model. Many of these strategies are now based on \acp{dnn}, which have shown remarkable success in capturing the subtleties of complex processes~\cite{Goodfellow-et-al-2016}. When there is no characterization of the dynamics, one can train deep learning systems designed for processing time sequences, e.g., \acp{rnn}~\cite{chung2014empirical} and attention mechanisms~\cite{vaswani2017attention}, for state estimation in intractable environments~\cite{becker2019recurrent}. Yet, they do not incorporate domain knowledge such as structured \ac{ss} models in a principled manner, while requiring many trainable parameters and large data sets even for simple sequence models~\cite{zaheer2017latent} and lack the interpretability of \ac{mb} methods. It is also possible to combine \acp{dnn} with variational inference in the context of state space models, as in~\cite{krishnan2015deep, archer2015black, karl2016deep, krishnan2017structured, fraccaro2017disentangled}. This is done by casting the Bayesian inference task as the optimization of a parameterized posterior and maximizing an objective. However, the learning procedure tends to be complex and prone to approximation errors since these methods often rely on highly parameterized models. Furthermore, their applicability to use cases with a bounded delay on hardware-limited devices is limited.

An alternative \ac{dd} approach for state estimation in \ac{ss} models uses \acp{dnn}  to encode the observations into some latent space that is assumed to obey a simple \ac{ss} model, typically a linear Gaussian one. State estimation is then carried out based on the extracted features~\cite{haarnoja2016backprop, laufer2018hybrid, zhou2020kfnet,coskun2017long, rangapuram2018deep}, and can be followed by another \ac{dnn} decoder~\cite{fraccaro2017disentangled}. This form of \ac{dnn}-aided state estimation is intended to cope with complex and intractable observations models, e.g., when processing visual observations, while one should still know (or estimate) the state evolution. When the \ac{ss} model is known, \acp{dnn} can be applied to improve upon \ac{mb} inference, as done in~\cite{satorras2019combining}, where graph neural networks are used in parallel with \ac{mb} smoothing. However, while the approach suggested in~\cite{satorras2019combining} uses \ac{dd} \acp{dnn}, it also requires full knowledge of the \ac{ss} model to apply \ac{mb} smoothing in parallel, as in traditional smoothers.

In scenarios involving partially known dynamics, where one has access to an approximation of some 
parts of the \ac{ss} model (based on, e.g., understanding of the underlying physics or established motion models), both \ac{mb} smoothing and \ac{dd} methods based on \acp{dnn} may be limited in their performance and suitability. In our previous work~\cite{KalmanNetTSP} we derived a hybrid \ac{mb}/\ac{dd} implementation of the \ac{kf} following the emerging \ac{mb} deep learning methodology~\cite{shlezinger2020model, shlezinger2022model,shlezinger2023model}. The augmentation of the \ac{kf} with a dedicated \ac{dnn} was shown to result in a filter that approaches \ac{mmse} performance in partially known dynamics, while being operable at high rates on limited hardware~\cite{LopezICAS21} and facilitating coping with high-dimensional observations~\cite{buchnik2023latent}. Further, the interpretable nature of the resulting architectures was leveraged to provide reliable measures of uncertainty~\cite{klein2021uncertainty} and support unsupervised training~\cite{revach2021unsupervised}. These findings, which all considered a filtering task, motivate deriving a hybrid \ac{mb}/\ac{dd} smoothing algorithm.

%
In this work, we introduce \acl{rn}, an iterative hybrid \ac{mb}/\ac{dd}  algorithm for smoothing in dynamical systems describable by partially known \ac{ss} models. \acl{rn} preserves the \ac{ks} flow, while converting it into a trainable machine learning architecture by replacing both the forward and backward \acp{kg} with compact \acp{rnn}. Additionally, \acl{rn} integrates an iterative refinement mechanism, enabling multiple iterations via deep unfolding~\cite{monga2021algorithm}. Consequently, \acl{rn} is able to convert a fixed number of \ac{ks} iterations into a discriminative model~\cite{shlezinger2022discriminative} that is trained \acl{e2e}.

\textcolor{NewColor}{
Although \acl{rn} learns the smoothing task from data, it preserves to flow of the \ac{ks}, thus retaining its recursive nature, low complexity, interpretability, and invariance to the length of the sequence.} In particular, \acl{rn} is shown to achieve the \ac{mmse} for linear models just as the \ac{ks} does with full information, while only having access to partial information, and notably outperforms the \ac{ks} when there is model mismatch. For \acl{nl} \ac{ss} models, \acl{rn} is shown to outperform \ac{mb} variants of the \ac{ks}, that are no longer optimal even with full domain knowledge. We also show that \acl{rn} outperforms leading \ac{dd} smoothers while using fewer trainable parameters, and being more efficient in terms of training and inference times.

The improved performance follows from the ability of \acl{rn} to follow the principled \ac{ks} operation, while circumventing its dependency on knowledge of the underlying noise statistics. In particular, by training the \acl{rn} smoother to directly compute the posterior distribution using learned \acp{kg}, we overcome the need to approximate the propagation of the noise statistics through the nonlinearity, \textcolor{NewColor}{and leverage data to cope with modeling mismatches}. Moreover, doing so also bypasses the need for numerically costly matrix inversions and linearizations required in the \ac{ks} equations.

The rest of this paper is organized as follows: 
\secref{sec:sysModel} reviews the \ac{ss} model and its associated tasks, and discusses relevant preliminaries. \secref{sec:RTSNet} details the discriminative architecture of \acl{rn}. \secref{sec:Results} presents the empirical study. \secref{sec:Conclusion} provides concluding remarks.

Throughout the paper, we use boldface lower-case letters for vectors and boldface upper-case letters for matrices. The transpose, $\ell_2$ norm, and stochastic expectation are denoted by $\set{\cdot}^\top$,  $\norm{\cdot}$, and  $\expecteds{\cdot}$, respectively. The Gaussian distribution with mean $\gmat{\mu}$ and covariance $\gmat{\Sigma}$ is denoted by $\mathcal{N}(\gmat{\mu}, \gmat{\Sigma})$. Finally, $\greal$ and $\gint$ are the sets of real and integer numbers, respectively.
%
%
%
\section{System Model and Preliminaries}\label{sec:sysModel}
%
%
\subsection{Problem Formulation}\label{ssec:SSModel}
{\bf \ac{ss} Models:} Dynamical systems in \acl{dt} describe the relationship between a sequence of observations $\gvec{y}_t$ and a sequence of unknown latent state variables $\gvec{x}_t$, where $t\in\gint$ is the time index. \ac{ss} models  are a common characterization of dynamic systems~\cite{bar2004estimation}, 
which in the (possibly) \acl{nl} and Gaussian case, take the form
\begin{subequations}\label{eq:NL_SS_model}
\begin{align}\label{eqn:stateEvolution}
\gvec{x}_{t}&= 
\gvec{f}\brackets{\gvec{x}_{t-1}}+\gvec{e}_t,
&\gvec{e}_t\sim
\mathcal{N}\brackets{\gvec{0},\gvec{Q}},
\hspace{0.25cm}
&\gvec{x}_{t}\in\greal^m,\\ \label{eqn:stateObservation}
\gvec{y}_{t}&=
\gvec{h}\brackets{\gvec{x}_{t}}+\gvec{v}_{t},
&\gvec{v}_t\sim
\mathcal{N}\brackets{\gvec{0},\gvec{R}},
\hspace{0.25cm}
&\gvec{y}_{t}\in\greal^n.    
\end{align}
\end{subequations}
In \eqref{eqn:stateEvolution}, the state vector $\gvec{x}_{t}$  evolves from the previous state $\gvec{x}_{t-1}$, by a (possibly) \acl{nl}, state-evolution function $\gvec{f}\brackets{\cdot}$ and by an \ac{awgn} $\gvec{e}_t$ with  covariance matrix $\gvec{Q}$. The observations $\gvec{y}_{t}$ in \eqref{eqn:stateObservation} are related to the current latent state vector by a (possibly) \acl{nl} observation mapping $\gvec{h}\brackets{\cdot}$ corrupted by \ac{awgn} $\gvec{v}_t$ with covariance $\gvec{R}$. A common special case of \eqref{eq:NL_SS_model} is that of linear Gaussian \ac{ss} models, where  there  exist matrices $\gvec{F},\gvec{H}$ such that
\begin{equation}\label{eqn:LinearSS}
\gvec{f}\brackets{\gvec{x}_{t-1}}={\gvec{F}}\cdot\gvec{x}_{t-1},
\quad
\gvec{h}\brackets{\gvec{x}_{t}}={\gvec{H}}\cdot\gvec{x}_{t}.   
\end{equation}

\smallskip
{\bf Smoothing Task:}
\ac{ss} models as in \eqref{eq:NL_SS_model} are studied in the context of several different tasks, which can be roughly classified into two main categories: observation recovery and hidden state estimation. The first category deals with recovering parts of the observed signal $\gvec{y}_t$. This corresponds, for example, to prediction and imputation. The second category lies at the core of the family of tracking problems, considering the estimation of  $\gvec{x}_t$. These include online (\acl{rt}) recovery, typically referred to as {\em filtering}, which is the task considered in~\cite{KalmanNetTSP}, and \emph{offline} estimation, i.e., \emph{smoothing}, which is the main focus of this paper. 
More specifically, smoothing involves the joint computation the state estimates $\hat{\gvec{x}}_{t\given{T}}$ in a given time horizon $T$, i.e., jointly estimating $\set{\gvec{x}_{t}}$ for each $t\in\set{1,2,\ldots,T} \triangleq \mathcal{T}$, given the corresponding block of noisy observations $\set{\gvec{y}_1,\gvec{y}_2,\ldots,\gvec{y}_T}$. 
%
%

\smallskip
{\bf Data-Aided Smoothing for Partially Known SS Models:} 
In practice, the \ac{ss} model parameters may be partially known, and one is likely to only have access to an approximated characterization of the underlying dynamics. We thus focus on such scenarios where the state-evolution function $\gevol$ and the state-observation function $\gobs$ can be reasonably approximated (possibly with mismatch) from our understating of the system dynamics and its physical design, or learned from data (as discussed in Subsection~\ref{subsec:discussion}). Regardless of how these functions are obtained, they can be used for smoothing. As opposed to the classical assumptions of the \ac{ks} algorithms,  the statistics of noises $\gvec{e}_t$ and $\gvec{v}_t$ are completely unknown, and may be non-Gaussian. 

To deal with the partial modeling of the dynamics, we assume access to a labeled \acl{ds} containing a sequence of observations and their corresponding states. Such data can be acquired, e.g., from field experiments, or using computationally intensive physically-compliant simulations~\cite{shlezinger2022model}. 
The \acl{ds} 
is comprised of $N$ time sequence pairs, i.e.,
%
$\mathcal{D}=\big\{({\gvec{Y}^{\brackets{i}}, \gvec{X}^{\brackets{i}}})\big\}_{i=1}^{N}$,
%
each of length $T_i$, namely,
\begin{equation}
\gvec{Y}^{\brackets{i}}=\sbrackets{
\gvec{y}_1^{\brackets{i}},\ldots,\gvec{y}_{T_i}^{\brackets{i}}}\in\greal^{n\times{T_i}}
\end{equation}
are the noisy observations, and the corresponding states are
\begin{equation}
\gvec{X}^{\brackets{i}}=\sbrackets{
\gvec{x}_0^{\brackets{i}},\gvec{x}_1^{\brackets{i}},\ldots,\gvec{x}_{T_i}^{\brackets{i}}}\in\greal^{m\times{T_i+1}}.
\end{equation}
Given $\mathcal{D}$ and the (approximated) $\gevol, \gobs$, our objective is to design a smoothing function which maps the observations $\{\gvec{y}_t\}_{t\in\mathcal{T}}$ into a state estimate $\{\hat{\gvec{x}}_t\}_{t\in\mathcal{T}}$, where the accuracy of the smoother is evaluated as the \ac{mse} with respect to the true state  $\{{\gvec{x}}_t\}_{t\in\mathcal{T}}$.
%
%
%
%
\subsection{Model-Based Kalman Smoothing}\label{ssec:MBKS}
We next recall the \ac{mb} \ac{rts} smoother~\cite{rauch1965maximum}, which is the basis for our proposed \acl{rn}, detailed in \secref{sec:RTSNet}. We describe the original algorithm for linear \ac{ss} models, as in \eqref{eqn:LinearSS}, and then discuss how to extend it for \acl{nl} \ac{ss} models. 

The \ac{rts} smoother recovers the latent state variables using two linear recursive steps, referred to as the \emph{forward} and \emph{backward} passes.  The forward pass is a standard \ac{kf}, while the backward pass recursively computes corrections to the forward estimate, based on future observations.
%

%
\smallskip
\textbf{Forward Pass:} 
The \ac{kf} produces a new estimate $\hat{\gvec{x}}_{t\given{t}}$ using its previous estimate $\hat{\gvec{x}}_{t-1\given{t-1}}$ 
and the observation $\gvec{y}_{t}$. 
For each $t\in\mathcal{T}$, the \ac{kf} operates in two steps:  \emph{prediction} and \emph{update}. 

The first step {predicts} the current \textit{a priori} statistical moments based on the previous \textit{a posteriori} moments. The moments of $\gvec{x}_t$ are computed using the knowledge of the evolution matrix $\gvec{F}$ as
\begin{subequations}\label{eqn:predict_evol}
\begin{align}\label{eqn:evol_1}
\hat{\gvec{x}}_{t\given{t-1}} &= 
\gvec{F}\cdot{\hat{\gvec{x}}_{t-1\given{t-1}}},\\\label{eqn:evol_2}
\mySigma_{t\given{t-1}} &=
{\gvec{F}}\cdot\mySigma_{t-1\given{t-1}}\cdot{\gvec{F}}^\top+\gvec{Q},
\end{align}
\end{subequations}
and the moments of the observations $\gvec{y}_t$ are computed based on the knowledge of the observation matrix $\gvec{H}$ as
\begin{subequations}\label{eqn:predict_obs}
\begin{align}\label{eqn:obs_1}
\hat{\gvec{y}}_{t\given{t-1}} &=
\gvec{H}\cdot\hat{\gvec{x}}_{t\given{t-1}},\\\label{eqn:obs_2}
{\gvec{S}}_{t\given{t-1}} &=
{\gvec{H}}\cdot\mySigma_{t\given{t-1}}\cdot{\gvec{H}}^\top+\gvec{R}.
\end{align}
\end{subequations}

In the {update} step, the \textit{a posteriori} state moments are computed based on the \textit{a priori} moments as
\begin{subequations}\label{eqn:update}
\begin{align}\label{eqn:update1}
\hat{\gvec{x}}_{t\given{t}}&=
\hat{\gvec{x}}_{t\given{t-1}}+\Kgain_{t}\cdot\Delta\gvec{y}_t,\\\label{eqn:update2}
{\mySigma}_{t\given{t}}&=
{\mySigma}_{t\given{t-1}}-\Kgain_{t}\cdot{\mathbf{S}}_{t\given{t-1}}\cdot\Kgain^{\top}_{t}.
\end{align}
\end{subequations}
Here, $\Kgain_{t}$ is the \ac{kg}, and it is given by
\begin{equation}\label{eq:FWGain}
\Kgain_{t}={\mySigma}_{t\given{t-1}}\cdot{\gvec{H}}^\top\cdot{\gvec{S}}^{-1}_{t\given{t-1}},
\end{equation}
while $\Delta\gvec{y}_t =\gvec{y}_t-\hat{\gvec{y}}_{t\given{t-1}}$ is the innovation, and is the only term that depends on the observed data.

\smallskip
\textbf{Backward pass:} The backward pass is similar in its structure to the update step in the \ac{kf}. For each $t\in\set{T-1,\ldots,1}$, the forward belief is corrected with future estimates via
\begin{subequations}\label{eq:BW_update}
\begin{align}\label{eq:BW_update1}
\hat{\gvec{x}}_{t\given{T}}&=
\hat{\gvec{x}}_{t\given{t}}+\Sgain_t\cdot \overleftarrow{\Delta} \gvec{x}_{t+1},\\\label{BW_update2}
{\mySigma}_{t\given{T}}&={\mySigma}_{t|t} - \Sgain_t\cdot \Delta{\mySigma}_{t+1\given{T}} \cdot \Sgain_t^{\top}.
\end{align}
\end{subequations}
Here, $\Sgain_t$ is the {backward} \ac{kg}, computed based on second-order statistical moments from the forward pass as 
\begin{equation}\label{eq:bw_gain}
\Sgain_t={\mySigma}_{t|t} \cdot {{\gvec{F}}}^{\top}\cdot{\mySigma}_{t+1\given{t}}^{-1}.
\end{equation}
The {difference} terms are given by
$\overleftarrow{\Delta}\gvec{x}_{t+1}=
\hat{\gvec{x}}_{t+1\given{T}} -  \hat{\gvec{x}}_{t+1\given{t}}$ and  $\Delta{\mySigma}_{t+1} = 
{\mySigma}_{t+1}-{\mySigma}_{t+1\given{t}}$. The \ac{ks} is \ac{mmse} optimal for linear Gaussian \ac{ss} models.

%
%
{\bf Extension to Nonlinear Dynamics:} 
 For \acl{nl} \ac{ss} models as in \eqref{eq:NL_SS_model},  the first-order statistical moments \eqref{eqn:evol_1} and \eqref{eqn:obs_1} are replaced with
\begin{subequations}\label{eqn:predict_NL}
\begin{align}\label{eqn:evol_NL}
\hat{\gvec{x}}_{t\given{t-1}} &= 
\gvec{f}\brackets{\hat{\gvec{x}}_{t-1}},\\\label{eqn:obs_NL}
\hat{\gvec{y}}_{t\given{t-1}} &=
\gvec{h}\brackets{\hat{\gvec{x}}_{t\given{t-1}}},
\end{align}
\end{subequations}
respectively. Unfortunately, the second-order moments cannot be propagated directly through the nonlinearity and thus must be approximated, resulting in methods that no longer share the \ac{mse} \textcolor{NewColor}{optimality} achieved in linear models. 

Among the methods proposed to approximate the second-order moments are the \acl{urts}, which is based on unscented transformations~\cite{sarkka2008unscented}, and \acp{ps} which use sequential sampling~\cite{gordon1993novel}. Arguably the most common \acl{nl} smoother is the \ac{eks}, which uses straightforward linearization. Specifically, when $\gvec{f}\brackets{\cdot}$ and $\gvec{h}\brackets{\cdot}$ are differentiable, the \ac{eks} linearizes them in a time-dependent manner. This is done using their partial derivative matrices (Jacobians), evaluated at $\hat{\gvec{x}}_{t-1\given{t-1}}$ and $\hat{\gvec{x}}_{t\given{t-1}}$, namely,
\begin{subequations}\label{eqn:Jacob}
\begin{align}
\hat{\gvec{F}}_t&=\jacob_f\brackets{\hat{\gvec{x}}_{t-1\given{t-1}}},\\
\hat{\gvec{H}}_t&=\jacob_h\brackets{\hat{\gvec{x}}_{t\given{t-1}}}.
\end{align}
\end{subequations} 
The Jabobians in \eqref{eqn:Jacob} are then substituted into equations \eqref{eqn:evol_2},  \eqref{eq:bw_gain}, \eqref{eqn:obs_2}, and \eqref{eq:FWGain} of the \ac{ks}. 

The forward and backward \acp{kg} are {pivot} terms, that are used as tuning factors for updating our current belief, and they depend on the second-order moments. For linear \ac{ss} models, the covariance computation is purely \ac{mb}, i.e., based solely on the noise statistics, while for \acl{nl} systems the covariance depends on the specific trajectory. 
Furthermore, these covariance computations require full knowledge of the underlying model, and performance notably degrades in the presence of a model mismatch. This motivates the derivation of a data-aided smoothing algorithm that estimates the \acp{kg} directly as a form of discriminative learning~\cite{shlezinger2022discriminative}, and by that circumvents the need to estimate the second-order moments. 
%
%
\section{RTSNet}\label{sec:RTSNet}
Here, we present \acl{rn}. We begin by describing our design rationale and high-level architecture in Subsection~\ref{subsec:Highlevel}, after which we detail the microarchitecture in Subsection~\ref{subsec:micro}. We then describe the training procedure in Subsection~\ref{subsec:training}, and provide a discussion in Subsection~\ref{subsec:discussion}.
%
%
\subsection{High Level Design}\label{subsec:Highlevel}
{\bf Rationale:}
The basic design idea behind the proposed \acl{rn} is to utilize the skeleton of the \ac{mb} \ac{rts} smoother, hence the name \acl{rn}, and to replace modules depending on unavailable domain knowledge, with trainable \acp{dnn}. By doing so, we convert the \ac{ks} into a discriminative algorithm, that can be trained in a supervised \acl{e2e} manner. 

Our design is based on two main guiding properties:
\begin{enumerate}[label={\em P\arabic*}]  
\item \label{itm:OneFold} The \ac{rts} operation, comprised of a single forward-backward pass, is not necessarily \ac{mmse} optimal when the \ac{ss} model is not linear Gaussian.
\item \label{itm:KGs} The \ac{rts} smoother requires the missing domain knowledge (i.e., the noise statistics) and linearization operations solely for computing the \acp{kg} in \eqref{eq:FWGain} and \eqref{eq:bw_gain}. 
\end{enumerate}

\smallskip
{\bf Unfolded Architecture (by \ref{itm:OneFold}):}
Property~\ref{itm:OneFold} indicates that one can possibly improve performance by carrying out multiple forward-backward passes, iteratively refining the estimated sequence. Following the deep unfolding methodology~\cite{shlezinger2022model}, we design \acl{rn} to carry out $K$ forward-backward passes, for some fixed $K\geq 1$.

Each pass of index $k \in \{1,\ldots, K\}$ involves a single forward-backward smoothing. The input and output of this pass are the sequences $\gvec{Y}_k=\sbrackets{\gvec{y}_{k,1},\gvec{y}_{k,2},\ldots,\gvec{y}_{k,T}}$ and 
${\hat{\gvec{X}}_k=\sbrackets{\hat{\gvec{x}}_{k,1\given{T}},\hat{\gvec{x}}_{k,2\given{T}},\ldots\hat{\gvec{x}}_{k,T\given{T}}}}$, respectively, and its operation is based on an \ac{ss} model of the form
\begin{subequations}\label{eq:Unfolded_SS_model}
\begin{align}\label{eqn:UnfoldedstateEvolution}
\gvec{x}_{k,t}&= 
\gvec{f}\brackets{\gvec{x}_{k,t-1}}+\gvec{e}_{k,t},
&\gvec{x}_{k,t}\in\greal^m,\\ \label{eqn:UnfoldedstateObservation}
\gvec{y}_{k,t}&=
\gvec{h}_k\brackets{\gvec{x}_{k,t}}+\gvec{v}_{k,t},
&\gvec{y}_{k,t}\in\greal^{n_k}.    
\end{align}
\end{subequations}
For the first pass, the inputs are $\gvec{y}_{1,t} =\gvec{y}_t$, i.e., the observations, and thus $\gvec{h}_1(\cdot) \equiv \gvec{h}(\cdot)$ and $n_1 = n$ in \eqref{eqn:UnfoldedstateObservation}. For the following passes where $k> 1$, the input is the estimate produced by the subsequent pass, i.e., 
${\gvec{y}_{k,t} =\hat{\gvec{x}}_{k-1,t\given{T}}}$. This input is treated as noisy state observations, and thus $\gvec{h}_{k}(\cdot) $ is the identity mapping and $n_k = m$. The noise signals in \eqref{eq:Unfolded_SS_model} obey an unknown distribution, following the problem formulation in Subsection~\ref{ssec:SSModel}.

\smallskip
{\bf Deep Augmenting \ac{rts} (by \ref{itm:KGs}):} 
We choose the \ac{rts} smoother as our \ac{mb} backbone based on \ref{itm:KGs}. Specifically,  as opposed to other alternatives, e.g., \acs{mbf}~\cite{bierman1977factorization, bierman1973fixed} and \acs{bifm}~\cite{wadehn2019state}, in \ac{rts} all the unknown domain knowledge  is encapsulated in the forward and backward \acp{kg}, $\Kgain_t$, and $\Sgain_t$, respectively. Consequently, for each pass $k$, we employ an \ac{rts} smoother, while replacing the \acp{kg} computation with  \acp{dnn}.  

Since both \acp{kg} involve tracking time-evolving second-order moments, they are replaced by \acp{rnn} in each pass of \acl{rn}, with input features encapsulating the missing statistics. The resulting operation of the $k$th pass commences with a forward pass, that is based upon  \acl{kn}~\cite{KalmanNetTSP}: For each $t$ from $1$ to $T$ a \emph{prediction} and \emph{update} steps are applied.
In the \emph{prediction} step, we use the prior estimates for the current state and for the current observation, as in \eqref{eqn:predict_NL}, namely, 
\begin{equation}
\hat{\gvec{x}}_{k,t|t-1} = \gvec{g}\brackets{\hat{\gvec{x}}_{k,t-1}}, \quad 
\hat{\gvec{y}}_{k,t|t-1} = \gvec{h}_k\brackets{\hat{\gvec{x}}_{k,t|t-1}}.   
\end{equation}
In the \emph{update} step, we compute $\FWpost{k,t}$, 
the current forward posterior, using \eqref{eqn:update1}, i.e., 
\begin{equation}
\hat{\gvec{x}}_{k,t\given{t}}=
\hat{\gvec{x}}_{k,t\given{t-1}}+\Kgain_{k,t}\cdot\brackets{\gvec{y}_{k,t}-\hat{\gvec{y}}_{k,t|t-1}}.
\end{equation}
As opposed to the \ac{ks}, here the filtering (forward) \ac{kg} $\Kgain_{k,t}$ is computed using an \ac{rnn}. 

\begin{figure*}
\begin{center}
%
%
\begin{subfigure}[pt]{0.99\columnwidth}
\includegraphics[width=1\columnwidth]{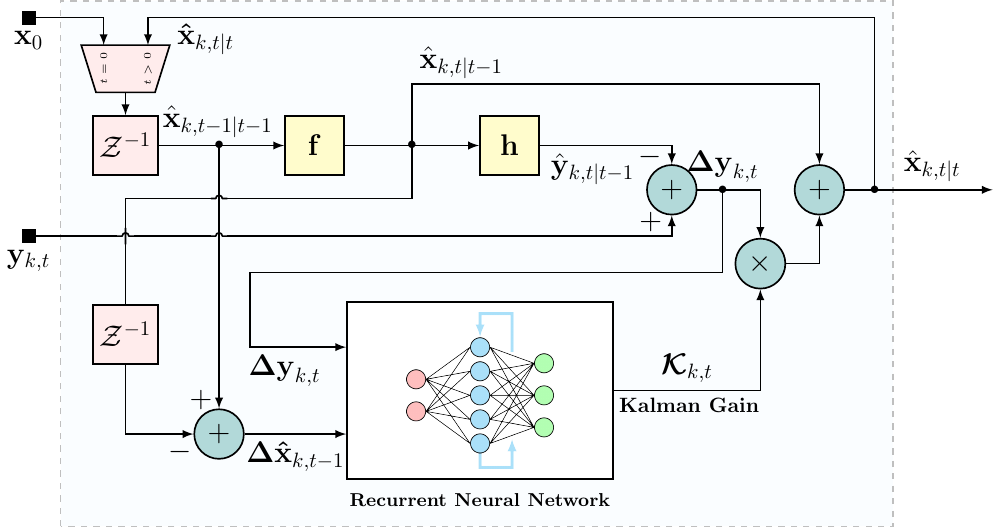}
\caption{$k^{th}$ forward pass.}
\label{fig:KNet1}
\end{subfigure}
%
%
\begin{subfigure}[pt]{0.99\columnwidth}
\vspace{0.35cm}
\includegraphics[width=1\columnwidth]{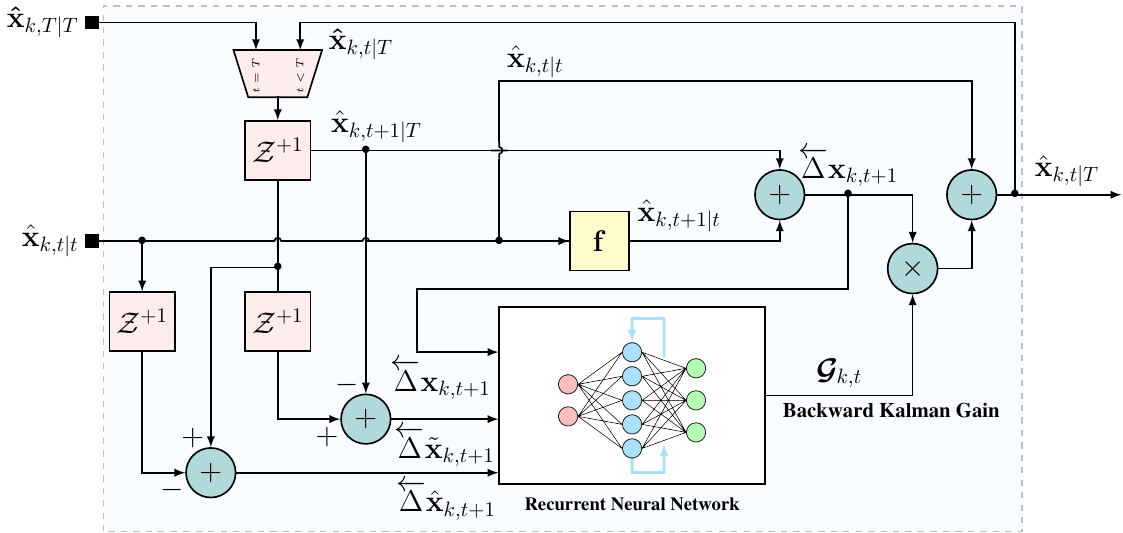}
\caption{$k^{th}$ backward pass.}
\label{fig:BWPass}
\end{subfigure}
%
%
\begin{subfigure}[pt]{2.5\columnwidth}
\vspace{0.25cm}
\hspace*{-1.5cm}  
\includegraphics[width=1\columnwidth]{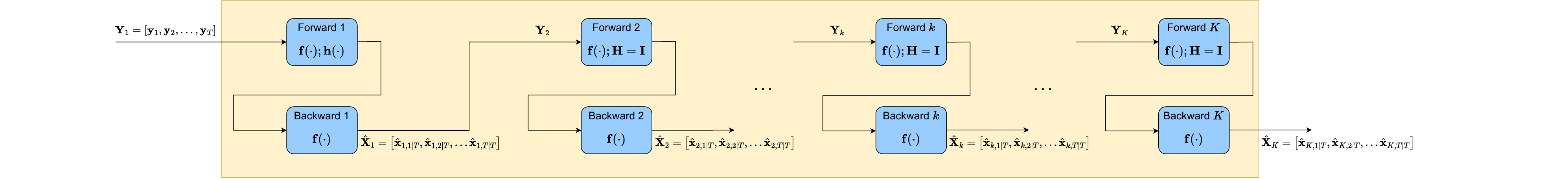}
\label{fig:Overall}
\end{subfigure}
\caption{\acl{rn} high level architecture block diagram.}
\label{fig:RTSNet_Macro}
\end{center}
\figSpace
\end{figure*}

The forward pass if followed by a backward pass, which updates our state estimates using information from future estimates. As in \eqref{eq:BW_update1}, this procedure is given by 
\begin{equation} 
\hat{\gvec{x}}_{k,t\given{T}}=
\hat{\gvec{x}}_{k,t\given{t}}+\Sgain_{k,t}\cdot\brackets{\hat{\gvec{x}}_{k,t+1\given{T}} -  \hat{\gvec{x}}_{k,t+1\given{t}}}.
\end{equation}
where the resulting estimate is $\hat{\gvec{x}}_{k,t} = \hat{\gvec{x}}_{k,t\given{T}}$ for each $t$. As in the forward pass, the smoothing (backward) \ac{kg} $\Sgain_t$ is computed using an \ac{rnn}. The  high-level architecture of \acl{rn} is depicted in~\figref{fig:RTSNet_Macro}.

%
%

%
\subsection{Microarchitecture}\label{subsec:micro}
\acl{rn} includes $2K$ \acp{rnn}, as each $k$th pass utilizes one \ac{rnn} to compute the forward \ac{kg} and another \ac{rnn} to compute the backward \ac{kg}. In this subsection, we focus on a single pass of index $k$ and formulate the microarchitecture of its forward and backward \acp{rnn}, as well as which features the \acp{rnn} use to compute the \acp{kg}. 

\begin{figure}
\includegraphics[width=0.95\columnwidth]{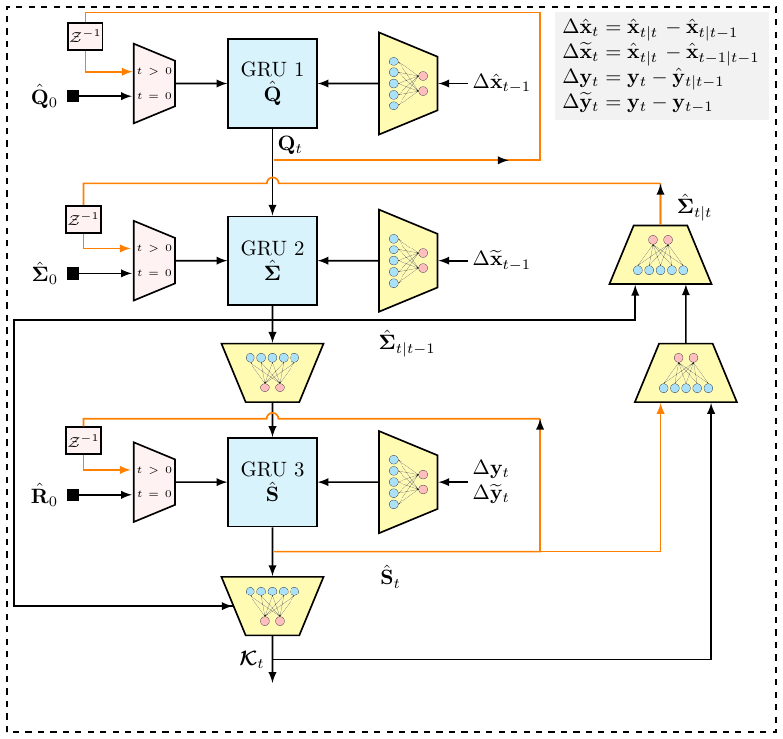}
\caption{Forward gain \ac{rnn} block diagram. The input features are used to update three \acp{gru} with dedicated \ac{fc} layers, and the overall interconnection between the blocks is based on the flow of the Forward \ac{kg} $\Kgain_t$ computation in the \ac{mb} \ac{kf}.}
\label{fig:FW_Gain_RNN} 
\end{figure}

\smallskip
{\bf Forward Gain:}
The forward pass is built on \acl{kn}, where architecture 2 of~\cite{KalmanNetTSP} is particularly utilized to compute the forward \ac{kg}, i.e., $\Kgain_{k,t}$, using separate \ac{gru} cells for each of the tracked \acl{cov}. The division of the architecture into separate \ac{gru} cells and \ac{fc} layers and their interconnection is illustrated in~\figref{fig:FW_Gain_RNN}. As shown in the figure, the network composes three \ac{gru} layers, connected in a cascade with dedicated input and output \ac{fc} layers.
This architecture, which is composed of a non-standard interconnection between \acp{gru} and \ac{fc} layers, is directly tailored towards the formulation of the \ac{ss} model and the operation of the \ac{mb} \ac{kf}, as detailed in~\cite{KalmanNetTSP}.

 The input features are designed to capture differences in the state and the observation model, as these differences are mostly affected by unknown noise statistics. As in~\cite{KalmanNetTSP}, the following features are  used to compute $\Kgain_{k,t}$ (see \figref{fig:FW_Gain_RNN}):
\begin{enumerate}[label={\em F\arabic*}]  
\item \label{itm:obDif}  \emph{Observation difference} $\Delta\tilde{\gvec{y}}_{k,t}={\gvec{y}_{k,t}-\gvec{y}_{k,t-1}}$.
\item \label{itm:inDif} \emph{Innovation difference} $\Delta\gvec{y}_{k,t}=\gvec{y}_{k,t}-\hat{\gvec{y}}_{k,t\given{t-1}}$.
\item \label{itm:FEDif} \emph{Forward evolution difference}  $\Delta\tilde{\gvec{x}}_{k,t}={\hat{\gvec{x}}_{k,t\given{t}}\!-\!\hat{\gvec{x}}_{k,{t\!-1}\given{t\!-1}}}$.
\item \label{itm:FUDif} \emph{Forward update difference}  $\Delta\hat{\gvec{x}}_{k,t}={\hat{\gvec{x}}_{k,t\given{t}}-\hat{\gvec{x}}_{k,{t}\given{t-1}}}$.
\end{enumerate}
%
%
%
%
%
\begin{figure} 
\includegraphics[width=0.95\columnwidth]{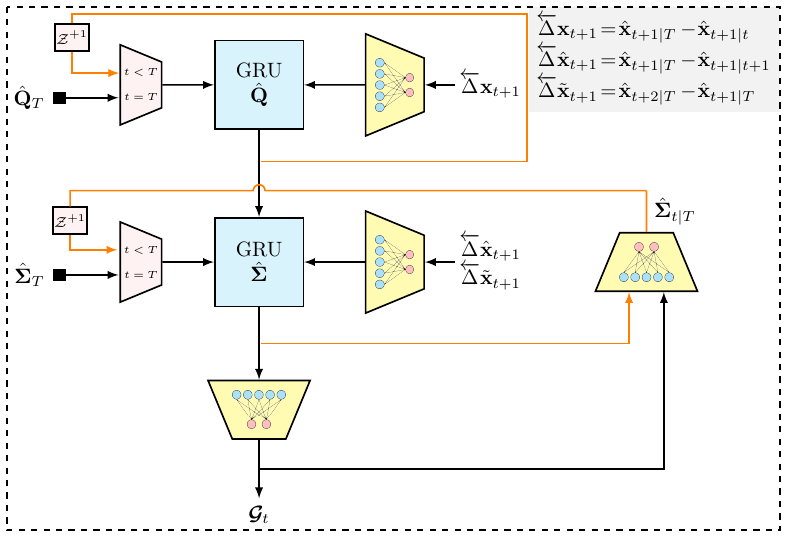}
\caption{{Backward gain \ac{rnn} block diagram. The input features are used to update two \acp{gru} with dedicated \ac{fc} layers, and the overall interconnection between the blocks is based on the flow of the Backward \ac{kg} $\Sgain_t$ computation in the \ac{ks}.}}
\label{fig:BW_Gain_RNN}
\end{figure}
%

%
%
\smallskip
{\bf Backward Gain:}
To compute $\Sgain_{k,t}$ in a learned manner, we again design an architecture based on how the \ac{ks} computes the backward gain. To that aim, we again use separate \ac{gru} cells for each of the tracked \acl{cov}, as illustrated in \figref{fig:BW_Gain_RNN}. The first \ac{gru} layer tracks the unknown state noise covariance $\gvec{Q}$, thus tracking $m^2$ variables. Similarly, the second  \acp{gru} tracks the predicted moment $\hat{\gvec{\Sigma}}_{t|T}$ \eqref{BW_update2} and $\hat{\gvec{S}}_t$ \eqref{eqn:obs_2}, thus having $m^2$  hidden state variables. The \acp{gru} are interconnected such that the learned $\gvec{Q}$ is used to compute $\hat{\gvec{\Sigma}}_{t|T}$, while \ac{fc} layers are utilized for input and output shaping.

We utilize the following features, which are related to the unknown underlying statistics:
\begin{enumerate}[label={\em B\arabic*}]  
\item {\em Update difference} $\overleftarrow{\Delta}\gvec{x}_{k,t+1}=
\hat{\gvec{x}}_{k,t+1\given{T}}-\hat{\gvec{x}}_{k,t+1\given{t}}$.
\item {\em Backward forward difference} $\overleftarrow{\Delta}\hat{\gvec{x}}_{k,t\!+1}\!=\! \\
\hat{\gvec{x}}_{k,t\!+\! 1\given{T}}\!-\!\hat{\gvec{x}}_{k,t\!+\! 1\given{t\!+\! 1}}$.
\item {\em Evolution difference} $\overleftarrow{\Delta}\tilde{\gvec{x}}_{k,t+1}=
\hat{\gvec{x}}_{k,t+2\given{T}}-\hat{\gvec{x}}_{k,t+1\given{T}}$.
\end{enumerate}
The first two features capture the uncertainty in the state estimate, where the differences remove predictable components such that they are mostly affected by the unknown noise statistics. The third feature is related to the evolution of the predicted state and thus reflects on its statistics that are tracked by the \ac{ks}. The features are utilized by the proposed architecture, as illustrated in~\figref{fig:BW_Gain_RNN}. 
%
%
\subsection{Training Algorithm}\label{subsec:training}
{\bf Loss Function:} 
We train \acl{rn} in a supervised manner. To formulate the loss function used for training, let $\NNParam_k$ denote the trainable parameters of the \acp{rnn} of the $k$th pass. The loss measure used to tune these parameters is the regularized $\ell_2$ loss; for a labeled pair $(\gvec{Y}_k{(i)}, \gvec{X}^{(i)})$ of length $T_i$, this loss is given by
\begin{align} 
\mathcal{L}_k^{(i)}\brackets{\NNParam_k} =&
\frac{1}{T_i}\cdot\sum_{t=1}^{T_i}
\norm{
\hat{\gvec{x}}_{k,t\given{T}}\brackets{\gvec{Y}_k^{(i)}, \NNParam_k}- \gvec{x}^{\brackets{i}}_t}^2 \notag \\
&+\gamma_k\cdot\norm{\NNParam_k}^2.
\label{eqn:LossFunc}
\end{align} 
In \eqref{eqn:LossFunc}, $\hat{\gvec{x}}_{k,t\given{T}}\brackets{\gvec{Y}_k, \NNParam_k} $ denotes the $t$th output of the $k$th pass with input $\gvec{Y}_k$ and \ac{rnn} parameters $\NNParam_k$. 
While the loss in \eqref{eqn:LossFunc} refers to a single trajectory $i$, we use it to optimize \acl{rn} using variants of mini-batch stochastic gradient descent. Here for every {batch} indexed by $j$, we choose $B < N$  trajectories indexed by $i_1^j, \ldots, i_B^j$, and compute the  mini-batch loss of the $k$th pass as
\begin{equation}
\bar{\mathcal{L}}_{k,j}\brackets{\NNParam_k}=
\frac{1}{B}\sum_{b=1}^B\mathcal{L}_k^{(i_b^j)}\brackets{\NNParam_k}.
\label{eqn:Loss}
\end{equation}

{\bf Gradient Computation:}
\acl{rn} uses the \acp{rnn} for computing the \acp{kg} rather than for directly producing the estimate $\hat{\gvec{x}}_{k,t|T}$. The loss function \eqref{eqn:LossFunc} \textcolor{NewColor}{enables the evaluation of} the overall system without having to externally provide ground truth values of the \acp{kg} for training purposes. 
Training \acl{rn} in \textcolor{NewColor}{an} \acl{e2e} manner thus builds upon the ability to {backpropagate} the loss to the computation of the \acp{kg}. One can obtain the loss gradient with respect to the \acp{kg} from the output of \acl{rn} since by combing \eqref{eqn:update1} and \eqref{eq:BW_update1} we get that
\begin{equation}
\hat{\gvec{x}}_{k,t\given{T}}=
\hat{\gvec{x}}_{k,t\given{t-1}}+\Kgain_{k,t}\cdot\Delta\gvec{y}_{k,t}+
\Sgain_{k,t}\cdot \overleftarrow{\Delta} \gvec{x}_{k,t+1}.
\end{equation}
Consequently, the gradients of the $\ell_2$ loss terms with respect to the \acp{kg} obey
\begin{subequations}
\begin{align}
\frac{\partial}{\partial\Kgain_{k,t}} \norm{\hat{\gvec{x}}_{k,t\given{T}}-\gvec{x}_{t}}^2
&\propto\brackets{\gvec{x}_{k,t\given{T}}-\gvec{x}_t}\cdot\ino{k,t}^\top\\
\frac{\partial}{\partial\Sgain_{k,t}}\norm{\hat{\gvec{x}}_{k,t\given{T}}-\gvec{x}_{t}}^2
&\propto
\brackets{\gvec{x}_{k,t\given{T}}-\gvec{x}_t}\cdot\overleftarrow{\Delta} \gvec{x}_{k,t+1}^\top,
\end{align}
\end{subequations}
which in turn allows to compute the gradient of the $\ell_2$ loss with respect to $\NNParam_k$ via the chain rule.
The gradient computation indicates that one can learn the computation of the \acp{kg} by training \acl{rn} \acl{e2e}.

{\bf End-to-End Training:}
The differentiable loss function in \eqref{eqn:LossFunc} allows \acl{e2e} training of a single forward-backward pass of index $k$. To train the overall unfolded \acl{rn}, we consider the following loss measures:

{\em Joint learning}, where the \acp{rnn} of all the passes are simultaneously using the labeled dataset $\mathcal{D}$. Here, we stack the trainable parameters as $\NNParam = \{\NNParam_k\}$, and set the loss function for the $i$th trajectory to
\color{NewColor}
\begin{align}
    \mathcal{L}^{(i)}\brackets{\NNParam} =& \sum_{k=1}^{K}\alpha_k \cdot \brackets{\frac{1}{T_i}\cdot\sum_{t=1}^{T_i}
\norm{
\hat{\gvec{x}}_{k,t\given{T}}\brackets{\gvec{Y}_1^{(i)}, \NNParam}- \gvec{x}^{\brackets{i}}_t}^2} \notag \\
&+\gamma\cdot\norm{\NNParam}^2.
\label{eqn:OverallLoss}
\end{align}
 where $\hat{\gvec{x}}_{k,t\given{T}}\brackets{\gvec{Y}_1, \NNParam} $ is the $t$th output of the $k$th forward-backward pass  when the input to \acl{rn} is $\gvec{Y}_1$ and its parameters are $\NNParam$. 
 \color{black}
 The coefficients $\{\alpha_k\}_{k=1}^K$ in \eqref{eqn:OverallLoss} balance the contribution of each pass to the loss -- setting $\alpha_k =0$ for $k<K$ evaluates \acl{rn} based solely on its output, while setting $\alpha_k \neq 0$ for $k<K$ encourages also the intermediate passes to provide accurate estimates. The ability to evaluate \acl{rn} during training, not just based on its output but also on its intermediate pass (i.e., with$\alpha_k \neq 0$ for $k<K$), is a direct outcome of its interpretable deep unfolded design. In conventional black-box \acp{dnn}, one typically cannot associate its internal features with an operational meaning and thus trains it solely based on its output. In contrast, our unfolded architecture ensures that internal modules produce a gradually refined estimate. A candidate setting is 
$\alpha_k=\log(1+k)$, see, e.g., \cite{samuel2019learning,lavi2023learn}.

{\em Sequential learning} repeats the training procedure $K$ times, training each pass of index $k$ after its preceding passes have been trained using the same dataset. 
Here, the dataset $\mathcal{D}$ is first used to train only $\NNParam_1$ using \eqref{eqn:LossFunc} with $k=1$; then, $\NNParam_1$ is frozen and $\NNParam_2$ is trained using \eqref{eqn:LossFunc} with $k=2$ and $\gvec{Y}_2^{(i)}=\hat{X}_1^{(i)}$, and the procedure repeats until $\NNParam_K$ is trained. 
This form of training, proposed in~\cite{shlezinger2019deepSIC}, exploits the modular structure of the unfolded architecture and tends to be data efficient and simpler to train compared with joint learning, and is also the form of training used in our empirical study presented in Section~\ref{sec:Results}. 
%
%
\subsection{Discussion}\label{subsec:discussion}
\acl{rn} is designed to operate in a hybrid \ac{dd}/\ac{mb} manner, combining \acl{dl} with the classical  \ac{ks}. It is designed by converting the \ac{eks} into a trainable architecture as a form of discriminative machine learning~\cite{shlezinger2022discriminative}, that directly learns the smoothing task from data while bypassing the need to carry out system identification~\cite{pillonetto2023deep}.
By identifying the specific noise-model-dependent computations of the \ac{ks} and replacing them with a dedicated \acp{rnn} integrated in the \ac{mb} flow, \acl{rn} benefits from the individual strengths of both \ac{dd} and \ac{mb} approaches.
We particularly note several core differences between \acl{rn} and its \ac{mb} counterpart. 
\textcolor{NewColor}{First, \acl{rn} is suitable for settings where full knowledge on an underlying   \ac{ss} model describing the dynamics is not available. Furthermore
Unlike the  \ac{eks}, \acl{rn} does not attempt to linearize the \ac{ss} model, and does not impose a statistical model on the noise signals, while avoiding the need to compute a Jacobian and invert a matrix at each iteration. This notably facilitates operation in high-dimensional \acl{nl} settings~\cite{buchnik2023latent}. In addition, \acl{rn} filters in a \acl{nl} manner, as, e.g., its forward \ac{kg} matrix depends on the input $\gvec{y}_t$. Moreover, \acl{rn} supports multiple learned forward-backward passes. While the number of passes is a hyperparameter that should be set based on system considerations and empirical evaluations, our numerical studies reported in \secref{sec:Results} show that using $K=2$ passes improves accuracy in \acl{nl} settings where the single pass \ac{eks} is not optimal.}
Due to these differences, \acl{rn} is more robust to model mismatch and can infer more efficiently compared with the \ac{ks}, as shown in \secref{sec:Results}.

Compared to purely \ac{dd} state estimation, \acl{rn} benefits from its model awareness, as it supports systematic inclusion of the available state evolution and observation functions, and does not have to learn its complete operation from data.  As empirically observed in \secref{sec:Results}, \acl{rn} achieves improved \ac{mse} compared to utilizing \acp{rnn} for \acl{e2e} state estimation, and also approaches the \ac{mmse} performance achieved by the \ac{ks} in linear Gaussian \ac{ss} models. 

Furthermore, the operation of \acl{rn} follows the flow of \ac{ks} rather than being utilized as a \acl{bb}. This implies that the intermediate features exchanged between its modules have a specific operation meaning, providing interpretability that is often scarce in \acl{e2e}, \acl{dl} systems. For instance, the fact that \acl{rn} learns to compute the \acp{kg} indicates the possibility of providing not only estimates of the state $\gvec{x}_t$, but also a measure of confidence in this estimate, as the \acp{kg} can be related to the covariance of the estimate, as initially explored for \acl{kn} in~\cite{klein2021uncertainty}. 

While \acl{rn} is inspired by \acl{kn}, and shares its architecture for the forward pass, the algorithms are fundamentally different. \acl{kn} is a filtering method, designed to operate in an adaptive sample-by-sample manner. \acl{rn} is a smoothing algorithm, which jointly processes a complete observed measurement sequence. The most notable difference is in the addition of a backward pass for \acl{rn} (which is the main extension of the \ac{ks} over the \ac{kf}). However, \acl{rn} also introduces joint learning along with the forward \ac{kg}; the ability to unfold the operation into multiple iterative forward-backward passes to facilitate coping with complex \acl{nl} dynamics; and a dedicated learning procedure, as detailed in Subsection~\ref{subsec:training}.

The fact that \acl{rn} preserves the \ac{ks} flow also indicates potential avenues for future research arising from its combination with established model-based extensions to the \acl{mb} \ac{ks}. One such avenue of future research involves the incorporation of adaptive priors~\cite{loeliger2016sparsity,wadehn2016outlier} or optimization-based extensions of the \ac{eks}~\cite{aravkin2017generalized,aravkin2011ell,aravkin2013sparse,aravkin2014optimization,aravkin2014smoothing} for tackling non-Gaussian distributions and outliers, while possibly leveraging emerging techniques for combining iterative methods with deep learning techniques~\cite{shlezinger2022model,agrawal2021learning}. An alternative research avenue involves extending \acl{rn} to constrained smoothing~\cite{amor2018constrained}. This can potentially be achieved by leveraging the fact that it preserves the operation of the \ac{eks} and thus supports complying with some forms of state constraints by principled incorporation of differentiable projection operators~\cite{liang2022ncvx}. These extensions of \acl{rn} are left for future investigation.

While we train \acl{rn} in a supervised manner using labeled data, the fact that it preserves the operation of the \ac{ks} indicates the possibility of training it in an \emph{unsupervised} manner using a loss function measuring consistency between its estimates and observations, e.g., $ \norm{\gvec{h}\brackets{\hat{\gvec{x}}_{t\given{T}}}-\gvec{y}_t}^2$. 
One can thus envision \acl{rn} being trained offline in a supervised manner, while tracking variations in the underlying \ac{ss} model at run-time using online self-supervision. This approach was initially explored for \acl{kn} in~\cite{revach2021unsupervised}, and we leave its extension to future work.
The ability to train in an unsupervised manner opens the door to use the backbone of \acl{rn} in more \ac{ss} related tasks other than smoothing, e.g., signal de-noising~\cite{locher2022hierarchical}, imputation, and prediction. Nonetheless, we leave the exploration of these extensions of \acl{rn} for future work.
%
%
\section{Experiments and Results}\label{sec:Results}
In this section, we present an extensive empirical study of \acl{rn}\footnote{The source code used in our empirical study along with the complete set of hyperparameters can be found at \url{https://github.com/KalmanNet/RTSNet_TSP}.}, evaluating its performance in multiple setups 
and comparing it with both \ac{mb} and \ac{dd} benchmark algorithms. We consider both linear Gaussian \ac{ss} models, where we identify the ability of \acl{rn} to coincide with the \ac{ks} (which is \ac{mse} optimal in such settings), as well as challenging \acl{nl} models. 

\subsection{Experimental Setup}\label{subsec:setup}
{\bf Smoothers:}
Our empirical study compares \acl{rn} with \ac{mb} and \ac{dd} counterparts. We use the \ac{ks} (\ac{rts} smoother) as the \ac{mb} benchmark for linear models. For \acl{nl} models, we use the \ac{eks} and the \ac{ps}~\cite{godsill2004monte}, where the latter is based on the forward-filter backward simulator of \cite{pyParticleEst} with $100$ particles and $10$ backward trajectories. The benchmark algorithms were optimized for performance by manually tuning the covariance matrices. This tuning is often essential to avoid divergence under model uncertainty as well as under dominant nonlinearities.

Our main \ac{dd} benchmark is the hybrid graph neural network-aided belief propagation smoother of~\cite{satorras2019combining} (referred to as \acl{hybrid}), which incorporates knowledge of the \ac{ss} model. We also compare with a \acl{bb} architecture using a \ac{vanilla}, which is comprised of bi-directional \ac{gru} with input and output \ac{fc} layers designed to have a similar number of trainable parameters as \acl{rn}. 
Both \ac{dnn}-aided benchmarks are empirically optimized and cross-validated to achieve their best training performance.

We use the term \emph{full information} to describe cases where $\gevol$ and $\gobs$  are accurately known. The term \emph{partial information} refers to the case where \acl{rn} and benchmark algorithms operate with some level of model {mismatch} in their available knowledge of the \ac{ss} model parameters. 
\acl{rn} and the \ac{dd} \ac{vanilla} operate without access to the noise covariance matrices (i.e., $\gvec{Q}$ and $\gvec{R}$), while their \ac{mb} counterparts operate with an accurate knowledge of the noise covariance matrices from which the data was generated. 

{\bf Evaluation:}
The metric used to evaluate the performance is the empirical mean and standard deviation of squared error, denoted,$\hat{\mu}$ and  $\hat{\sigma}$, respectively.  Unless stated otherwise, we evaluate using a $N_{\mathrm{test}}=200$ test trajectories. 

Throughout the empirical study and unless stated otherwise, in the experiments involving synthetic data, the \ac{ss} model is generated using diagonal noise covariance matrices; i.e., 
\begin{equation}
\label{eqn:nuDef}
\gvec{Q}=\mathrm{q}^2\cdot\gvec{I},
\quad
\gvec{R}=\mathrm{r}^2\cdot\gvec{I},
\quad
\nu\triangleq\frac{\mathrm{q}^2}{\mathrm{r}^2}. 
\end{equation}
\textcolor{NewColor}{
By \eqref{eqn:nuDef}, setting $\nu$ to be $0$ dB implies that both the state noise and the observation noise have the same variance.}

%
\begin{table}
\caption{Linear \ac{ss} model - Full Information - \ac{mse} $\dB$}
\begin{subtable}[]{0.99\columnwidth}
\caption{$\nu=0\dB$}
\vspace{-0.3cm}
\begin{center}
\scriptsize{
\begin{tabular}{|c|c|c|c|c|c|c|c| }
\hline
\rowcolor{lightgray}\multicolumn{1}{|c|}{$\gscal{r}^{2} \dB $} & $10$ & $0$ & $-10$ & $-20$ & $-30$ \\
\hline
\textrm{Noise} & 10.023 & 0.054 & -10.003 &	-19.947 &	-29.962
\\
& $\pm$ 0.424 & $\pm$ 	0.448 & $\pm$ 0.427 & $\pm$ 0.411 & $\pm$ 0.430
\\
\hline
\ac{kf} & 8.085 &	-1.827 & -11.880 & -21.903 & -31.886
\\
& $\pm$ 0.502 & $\pm$ 0.525 & $\pm$ 0.464 & $\pm$ 0.419 & $\pm$ 0.514
\\
\hline
\rowcolor{CornflowerBlue}
\ac{ks} & 6.215 & -3.710 & -13.776 & -23.751 & -33.749
\\
\rowcolor{CornflowerBlue}
& $\pm$ 0.487 & $\pm$ 0.535 & $\pm$ 0.466 & $\pm$ 0.519 & $\pm$ 0.508
\\
\hline
\rowcolor{SeaGreen}
\acl{rn} & 6.225 & -3.695 & -13.738 & -23.732 & -33.698	
\\
\rowcolor{SeaGreen}
& $\pm$  0.487	& $\pm$ 0.537  & $\pm$ 0.463  & $\pm$ 0.512  & $\pm$ 0.506	
\\
\hline
\end{tabular}
\label{tbl:Baseline_0}
}
\end{center}
\end{subtable}
\vspace{0.15cm}

\begin{subtable}[]{0.99\columnwidth}
\caption{$\nu=-10\dB$}
\vspace{-0.3cm}
\begin{center}
\scriptsize{
\begin{tabular}{|c|c|c|c|c|c|c|c| }
\hline
\rowcolor{lightgray}\multicolumn{1}{|c|}{$\gscal{r}^{2} \dB $} & $10$ & $0$ & $-10$ & $-20$ & $-30$ \\
\hline
\textrm{Noise} & 10.004 & 0.000 & -10.029 &	-19.982 &	-29.969
\\
& $\pm$ 0.419 & $\pm$ 0.392 & $\pm$ 0.431 & $\pm$ 0.404 & $\pm$ 0.435
\\
\hline
\ac{kf} & 5.299 &	-4.703 & -14.756 & -24.680 & -34.731
\\
& $\pm$ 0.710 & $\pm$ 0.663 & $\pm$ 0.675 & $\pm$ 0.596 & $\pm$ 0.696
\\
\hline
\rowcolor{CornflowerBlue}
\ac{ks} & 1.834 & -8.220 & -18.179 & -28.098 & -38.236
\\
\rowcolor{CornflowerBlue}
& $\pm$ 0.794 & $\pm$ 0.721 & $\pm$ 0.778 & $\pm$ 0.726 & $\pm$ 0.837
\\
\hline
\rowcolor{SeaGreen}
\acl{rn} & 1.881 & -8.169 & -18.092 & -27.875 & -38.183 	\\
\rowcolor{SeaGreen}
& $\pm$ 0.796 & $\pm$ 0.720 & $\pm$ 0.797 & $\pm$ 0.746 & $\pm$	0.836
\\
\hline
\end{tabular}
\label{tbl:Baseline_m10}
}
\end{center}
\end{subtable}
\vspace{0.15cm}

\begin{subtable}[]{0.99\columnwidth}
\caption{$\nu=-20\dB$}
\vspace{-0.3cm}
\begin{center}
\scriptsize{
\begin{tabular}{|c|c|c|c|c|c|c|c| }
\hline
\rowcolor{lightgray}\multicolumn{1}{|c|}{$\gscal{r}^{2} \dB $} & $10$ & $0$ & $-10$ & $-20$ & $-30$ \\
\hline
\textrm{Noise} & 10.012 & -0.008 & -10.004 & -19.977 & -30.003
\\
& $\pm$ 0.398 & $\pm$ 0.434 & $\pm$ 0.448 & $\pm$ 0.416 & $\pm$ 0.429
\\
\hline
\ac{kf} & 2.756 & -7.254 & -17.339 & -27.194 & -37.324
\\
& $\pm$ 1.086 & $\pm$ 1.012 & $\pm$ 0.967 & $\pm$ 1.011 & $\pm$ 0.963
\\
\hline
\rowcolor{CornflowerBlue}
\ac{ks} & -1.790 & -11.847 & -21.738 & -31.620 & -41.831
\\
\rowcolor{CornflowerBlue}
& $\pm$ 1.242	& $\pm$ 1.190 & $\pm$ 1.281 & $\pm$ 1.107	& $\pm$ 1.289
\\
\hline
\rowcolor{SeaGreen}
\acl{rn} & -1.640 & -11.712 & -21.543 & -31.817 &	-41.505
\\
\rowcolor{SeaGreen}
& $\pm$ 1.199 & $\pm$ 1.173 & $\pm$	1.279 & $\pm$ 1.152 & $\pm$ 1.229
\\
\hline
\end{tabular}
\label{tbl:Baseline_m20}
}
\end{center}
\end{subtable}
\label{tbl:baseline}
%
\end{table}
%
%
\begin{figure}
\includegraphics[width=1\columnwidth]{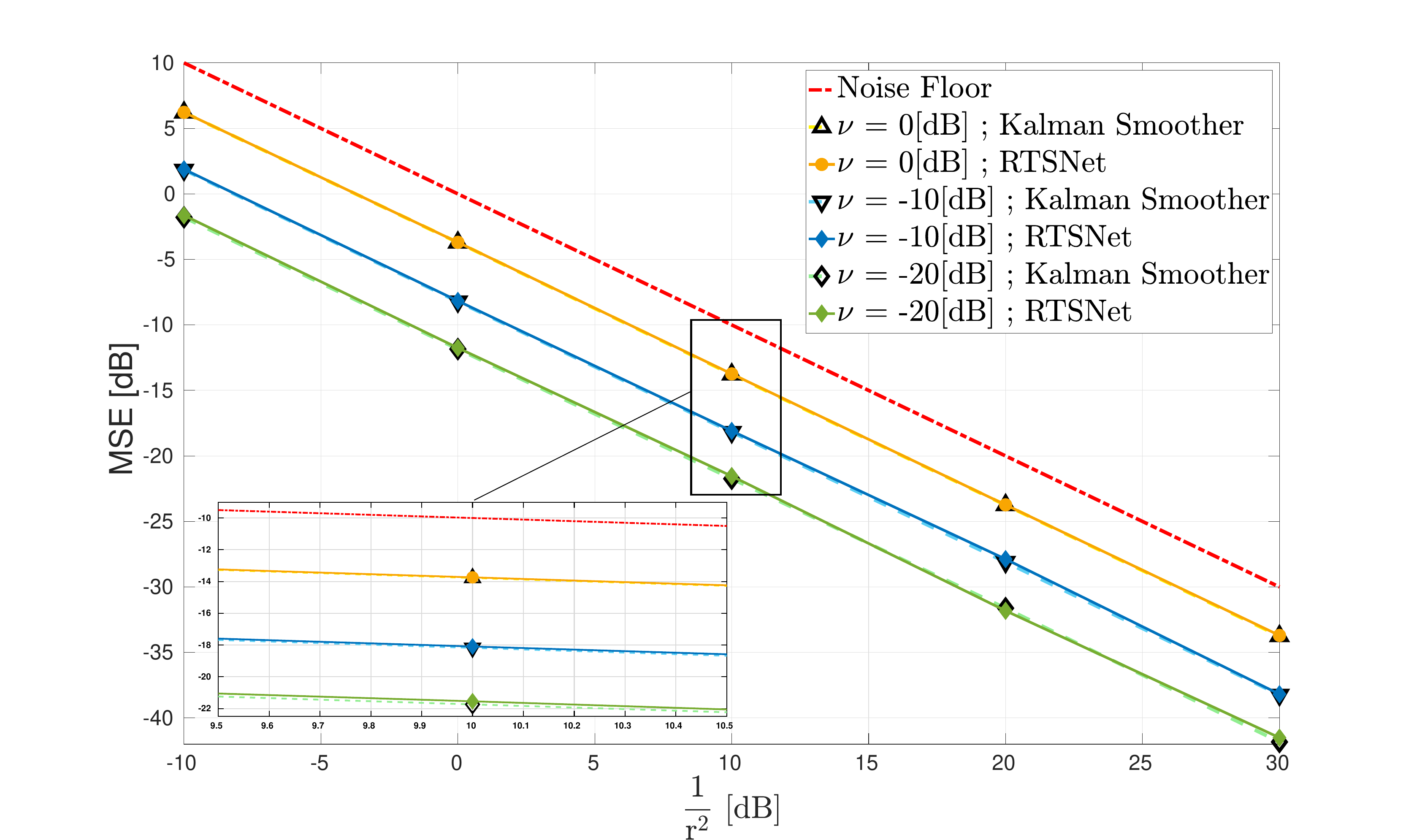}
\caption{Linear SS model - full information.}
\label{fig:Lin_Base} 
\end{figure}
%
%
\vspace{0.3cm}
\subsection{Linear State Space Models with Full Information}\label{subsec:SynLinear}
We first focus on comparing \acl{rn} with the \ac{ks} for synthetically generated {linear} Gaussian dynamics with full information. Since the \ac{ks} is \ac{mse} optimal here, we show that the performance of both algorithms coincides, demonstrating that \acl{rn} with $K=1$ can learn to be optimal. Then, we show that the learning capabilities of \acl{rn} are scalable, namely, that they hold for different \ac{ss} dimensions, and transferable, i.e., that they can be trained and evaluated with different trajectory lengths and initial conditions. \textcolor{NewColor}{We conclude this study by demonstrating the ability of \acl{rn} to learn to smooth in the presence of non-Gaussian \ac{ss} models.}

{\bf Approaching Optimality:}
We consider a $2\times2$ \ac{ss} model \eqref{eq:NL_SS_model}-\eqref{eqn:LinearSS},  where $\gvec{F}$ takes a \emph{canonical} form and $\gvec{H}$ is set to be the identity matrix, namely,
\begin{equation}\label{eq:F_canonical}
\gvec{F}=\begin{pmatrix}
1 & 1 \\
0 & 1
\end{pmatrix},
\quad
\gvec{H}=\gvec{I}.
\end{equation}
%
We use multiple noise levels, in $\dB$ scale, of
%
$\gscal{r^2}\in\set{10, 0, -10, -20, -30}$,
and 
$\nu\in\set{-20, -10, 0}$.
%
The results provided in \tbref{tbl:baseline}, and in \figref{fig:Lin_Base}, show that \acl{rn} converges to the \ac{mmse} estimate produced by the \ac{ks} in the first two moments. This indicates that \acl{rn} successfully learns to implement the \ac{ks} when it is \ac{mmse} optimal.

\smallskip
{\bf Scaling up Model Size:}
Next, we provide empirical evidence that \acl{rn} is a scalable smoothing architecture, capable of handling \ac{ss} models beyond just those with small dimensions. In this experiment $\gvec{F}$ and $\gvec{H}$ in their canonical form were considered, $\gscal{q^2}=-20\dB$, $\gscal{r^2}=0\dB$, and $T=20$. It is clearly observed in \tbref{tbl:lin_scaling} that \acl{rn} retains its optimality also for high dimensional models, outperforming its \ac{dd} benchmarks: \ac{vanilla} is far from optimal, and the performance of \acl{hybrid} degrades when the model dimensions increase.
%
%

\smallskip
{\bf Trajectory Length:}
To show generalization in $T$, the \ac{ss} model in \eqref{eq:F_canonical} is again considered, where $\gscal{q^2}=-20\dB$, and  $\gscal{r^2}=0\dB$. Here, we first train \acl{rn} and its \ac{dd} benchmarks on one trajectory length and then test it on a longer one. The results reported in \tbref{tbl:lin_gen_T} show that while \acl{rn} retains optimally for various trajectory lengths, \ac{vanilla} completely diverges. We can also see the superiority of \acl{rn} over \acl{hybrid} demonstrating slightly degraded performance. 

\smallskip
{\bf Initial Conditions:}
The operation of all smoothing algorithms depends on their initial state. We next train the \ac{dd} benchmarks on trajectories with a different initial state compared to that used in the test for the \ac{ss} model in \eqref{eq:F_canonical} with $\gscal{q^2}=-20\dB$, $\gscal{r^2}=0\dB$, and $T=100$. The results provided in \tbref{tbl:lin_gen_ic} demonstrates that 
while \ac{vanilla} completely diverges, and \acl{hybrid} is with slightly degraded performance when trained and then tested on different initial conditions, \acl{rn} still retains its optimally, which again demonstrates that it learns the smoothing task, rather than to overfit to trajectories presented during training.

\begin{figure}
\includegraphics[width=1\columnwidth]{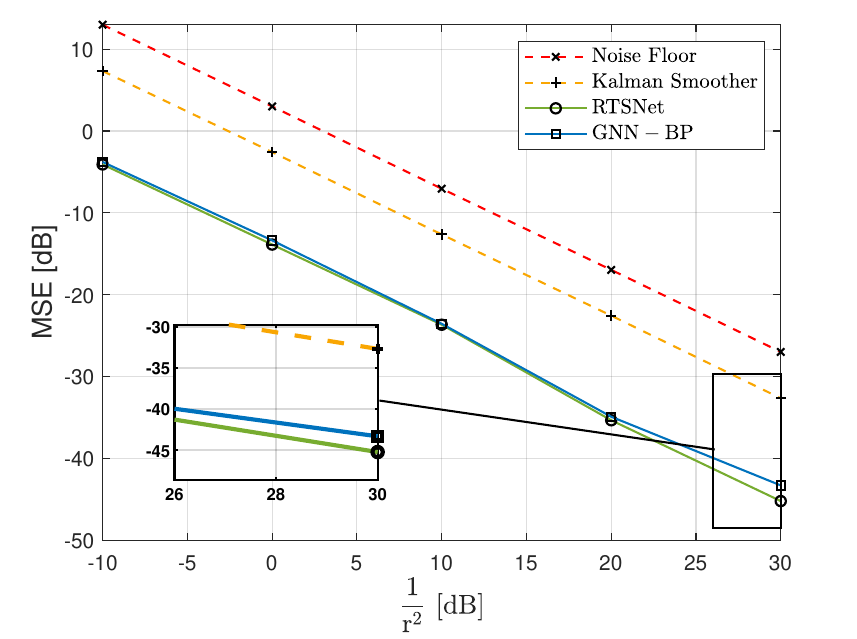}
\caption{Linear non-Gaussian SS model - full information.}
\label{fig:EXP_Noise} 
\end{figure}

\color{NewColor}
\smallskip
{\bf Non-Gaussian Noise:}
The fact that \acl{rn} augments the computation of the forward and backward gains with dedicated \acp{rnn} enables it to track in non-Gaussian dynamics, where the \ac{mb} \ac{ks} is no longer optimal. To demonstrate this, we consider a linear \ac{ss} model as in \eqref{eq:F_canonical} where the noise signals are drawn from an i.i.d. exponential distributions with covariance matrices given in \eqref{eqn:nuDef}. For this setting, we compare \acl{rn} with the \ac{mb} \ac{ks} and \ac{kf}, as well as with the \ac{dd} \acl{hybrid}, for $\nu = -20 \dB$. The resulting \ac{mse} versus $1/\gscal{r}^2$ are reported in~\figref{fig:EXP_Noise}. There, it is clearly observed that \acl{rn} outperforms not only the \ac{mb} \ac{ks} with a notable gap, but also the \ac{dd} \acl{hybrid}. These results indicate the ability of the hybrid architecture of \acl{rn} to successfully cope with non-Gaussian dynamics.

%

\color{black}
\begin{table}{}
\caption{Linear \ac{ss} model - Learning Capabilities}
%
%
\begin{subtable}[]{0.99\columnwidth}
\vspace{0.05cm}
\caption{Scaling \ac{ss} Model Dimensions}
\vspace{-0.3cm}
\begin{center}
\scriptsize{
\begin{tabular}{|c|c|c|c|c|c|c|}
\hline
\rowcolor{lightgray}
\multicolumn{1}{|c|}{Dimensions} & $2\times2$ & $5\times5$ & $10\times10$ & $20\times20$ \\
\hline
\ac{kf} &
-7.791	& -10.931 &	-11.062	& -11.540
\\
& $\pm$  2.204 & $\pm$ 1.411 & $\pm$ 0.951 & $\pm$ 0.771
\\
\hline
\rowcolor{CornflowerBlue}
\ac{ks} & -11.732	& -12.350 &	-12.436	&	-12.762
\\
\rowcolor{CornflowerBlue}
& $\pm$ 2.489 & $\pm$ 1.561 & $\pm$ 1.058 & $\pm$ 0.808
\\
\hline
\ac{vanilla} & 3.289	& 4.261	& 5.581	&	3.742
\\
& $\pm$ 4.495 & $\pm$ 4.724 & $\pm$ 4.553 & $\pm$ 4.416
\\
\hline
%
\acl{hybrid} & 
-10.016 & -9.028 & -8.674 & -8.557
\\
& $\pm$ 2.331 & $\pm$ 1.312 & $\pm$	0.803 & $\pm$ 0.603
\\
\hline
\rowcolor{SeaGreen}
\acl{rn} & -11.208 & -11.9725 & -12.0231 & -12.2755
\\
\rowcolor{SeaGreen}
& $\pm$ 2.438 & $\pm$ 1.597 & $\pm$ 1.055 & $\pm$ 0.828
\\
\hline
\end{tabular}
\label{tbl:lin_scaling}
}
\end{center}
\end{subtable}
%
%
\begin{subtable}[]{0.99\columnwidth}
\vspace{0.3cm}
\caption{Scalability for Trajectory Length} 
\vspace{-0.3cm}
\begin{center}
\scriptsize{
\begin{tabular}{|c|c|c|c|c|c|c|}
\hline
%
\rowcolor{lightgray}\multicolumn{1}{|c|}{$T_{\textrm{training}},T_{\textrm{testing}}$}
& $20, 20$ & $100, 100$ & $100, 1000$ & 
$100, \mathcal{U}\sbrackets{100,1000}$\\
\hline
Noise & 
0.025 & -0.008 & 0.011 & 0.015
\\
& $\pm$0.919 & $\pm$ 0.434 & $\pm$ 0.132 & $\pm$ 0.227
\\
\hline
\ac{kf} &
-7.791 & -7.254 & -7.162	& -7.241
\\
& $\pm$ 2.204 & $\pm$ 1.012 & $\pm$ 0.335 & $\pm$ 0.543
\\
\hline
\rowcolor{CornflowerBlue}
\ac{ks} &
-11.732 & -11.847 & -11.810 & -11.853
\\
\rowcolor{CornflowerBlue}
& $\pm$ 2.489 & $\pm$ 1.190 & $\pm$ 0.450 & $\pm$ 0.687
\\
\hline
\ac{vanilla} &
3.289 & 22.277 & 54.955 & 48.390
\\
& $\pm$ 4.495 & $\pm$ 5.190 & 4.419 & 5.436
\\
\hline
\acl{hybrid} &
 -10.016 & -11.433 & -11.662 & -11.687
\\
& $\pm$ 2.331 & $\pm$ 1.166 & $\pm$ 0.448 & $\pm$ 1.740
\\
\hline
\rowcolor{SeaGreen}
\acl{rn} &
-11.208 & -11.753 & -11.753 & -11.773
\\
\rowcolor{SeaGreen}
& $\pm$ 2.438 & $\pm$ 1.182 & $\pm$ 0.449 & $\pm$ 0.685
\\
\hline
\end{tabular}
\label{tbl:lin_gen_T}
}
\end{center}
\vspace{0.05cm}
\end{subtable}
%
%
\begin{subtable}[]{0.99\columnwidth}
\vspace{0.05cm}
\caption{Initial Conditions}
\vspace{-0.3cm}
\begin{center}
\scriptsize{
\begin{tabular}{|c|c|c|c|c|c|}
\hline
\rowcolor{lightgray}
\multicolumn{1}{|c|}{Training, Testing} & Fixed, Fixed & Fixed, Random & Random, Random  
\\
\hline
Noise & 
-0.008 & NA & -0.019
\\
& $\pm$ 0.434  & NA  & $\pm$ 0.360
\\
\hline
\ac{kf} & 
-7.254 & NA & -7.426
\\
& $\pm$ 1.012 & NA  & $\pm$ 0.963
\\
\hline
\rowcolor{CornflowerBlue}
\ac{ks} & 
-11.847	& NA & -12.025
\\
\rowcolor{CornflowerBlue}
& $\pm$ 1.190 & NA	&  $\pm$ 1.238
\\
\hline
\ac{vanilla} & 
22.277 & 37.281 & 26.606
\\
& $\pm$ 5.190 & $\pm$ 2.003 & $\pm$ 3.547
\\
\hline
\acl{hybrid} & 
-11.433 & -10.655 & -11.382
\\
& $\pm$ 1.166 & $\pm$ 1.219 & $\pm$ 1.164
\\
\hline
\rowcolor{SeaGreen}
\acl{rn} &
-11.753 & -11.757 & -11.701
\\
\rowcolor{SeaGreen}
& $\pm$ 1.182  & $\pm$ 1.187  & $\pm$ 1.214
\\
\hline
\end{tabular}
\label{tbl:lin_gen_ic}
}
\end{center}
\vspace{0.05cm}
\end{subtable}
%
\end{table}
%
%
\subsection{Linear SS Models with Partial Information}
\label{ssec:partialLinear}
%
Next, we demonstrate the merits of using \acl{rn} in linear settings with \emph{partial information} where the \ac{ks} is degraded due to the missing information. We consider mismatches in both the observation model as well as in the state evolution model. In particular, a mismatch in a model, e.g., the observation model, refers to a case in which a wrong setting of $\gvec{h}(\cdot)$ is used, while the \ac{mb} smoothers have access to the noise distribution. We also consider the case in which the corresponding model is unknown, e.g., both  $\gvec{h}(\cdot)$ and the noise distribution are unknown for the observation model; In such cases, since unlike \acl{hybrid} and the \ac{mb} benchmarks, \acl{rn} does not require prior knowledge of the noise distribution, we also evaluate it when it uses its data also to estimate the missing design parameter, e.g., $\gvec{h}(\cdot)$, using  \ac{ls}.

%
%

\smallskip
{\bf Observation Model Mismatch:}
We first consider the case where the \emph{design} observation model is mismatched. We again use the canonical model in \eqref{eq:F_canonical} with the observation matrix being 
either \emph{unknown}, or assumed to be $\gvec{H}_{0}=\gvec{I}$, while the observation model is
\begin{equation*}
\gvec{H}_{\alpha^\circ}=\gvec{R}_{x,y}\brackets{\alpha}\cdot\gvec{H}_{0}, \quad \gvec{R}_{x,y}\brackets{\alpha}=
\begin{pmatrix}
\cos{\alpha} & -\sin{\alpha} & 0  \\
\sin{\alpha} & \cos{\alpha} & 0 \\
0 & 0 & 1
\end{pmatrix},
\end{equation*}
%
%
with $\gvec{H}_{0}=\gvec{I}$. The data is generated with  $\alpha^\circ=10$. Such scenarios represent a setup where the true observed values are rotated by $\alpha^\circ$, e.g., a slight misalignment of the sensors exists. 

\begin{figure}
\vspace{-0.35cm}
\includegraphics[width=1\columnwidth]{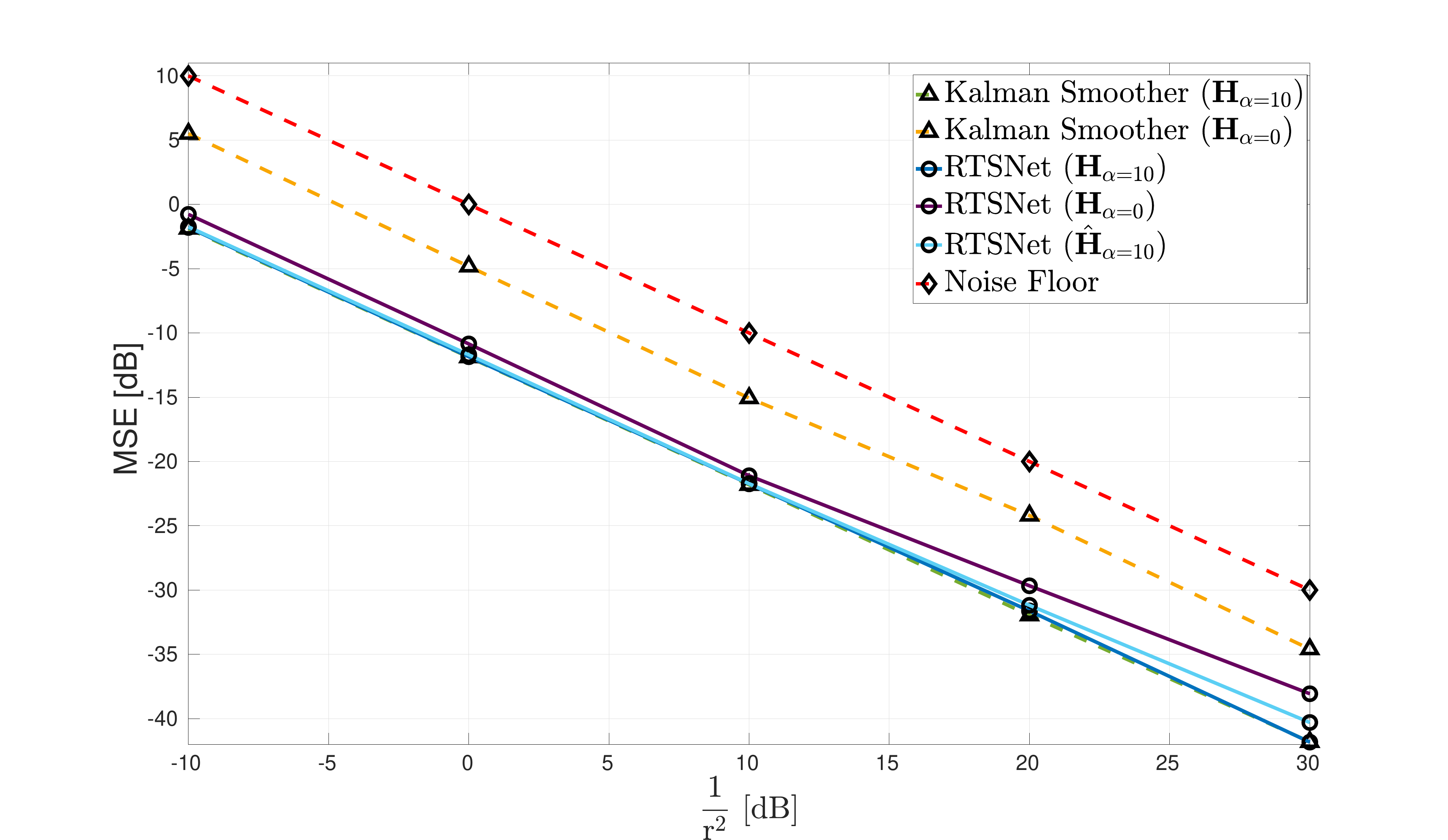}
\caption{Linear SS model - Observation mismatch.}
\label{fig:Lin_obs_mismatch}
\end{figure}

We compare the performance of \acl{rn} and the \ac{ks} when their design observation model is either $\gvec{H}_{\alpha=10}$ (full information) or $\gvec{H}_{\alpha=0}$ (mismatch). We also consider the case where the matrix is estimated from the data set via \ac{ls}, denoting  $\hat{\gvec{H}}_{\alpha=10}$. 
The empirical results reported in \tbref{tbl:lin_obs_miss} and in \figref{fig:Lin_obs_mismatch} demonstrate that while \ac{ks} experienced a severe performance degradation, \acl{rn} is able to compensate for mismatches using the learned \ac{kg}. When assuming a mismatched model $\gvec{H}_{0}$, \acl{rn} converges to within a minor gap from the \ac{mmse}, which is further reduced when the data is also used to estimate the observation model. The latter indicates that even when the \ac{ss} model is completely unknown, yet can be postulated as being linear, \acl{rn} can reliably smooth by using its data to both estimate the state evolution matrix as well as learn to smooth, while bypassing the need to impose a model on the noise.   

%
%
\begin{table}
\vspace{0.3cm}
\caption{Linear \ac{ss} - Observation mismatch: $\nu=-20\dB$}
\vspace{-0.3cm}
\begin{center}
\scriptsize{
\begin{tabular}{|c|c|c|c|c|c|c|c| }
\hline
\rowcolor{lightgray}\multicolumn{2}{|c|}{$\gscal{r}^{2} \dB $} & $10$ & $0$ & $-10$ & $-20$ & $-30$ \\
\hline
\ac{kf} & Full &
2.702 & -7.394 & -17.367 & -27.293 & -37.273
\\
&
& $\pm$ 0.885 & $\pm$ 0.901 & $\pm$	0.957 & $\pm$ 0.966	& $\pm$ 1.029
\\
\hline
\rowcolor{CornflowerBlue}
\ac{ks} & Full &
-1.875 & -11.880 & -21.812 & -31.961 & -41.810
\\
\rowcolor{CornflowerBlue}
&
& $\pm$ 1.285 & $\pm$ 1.272 & $\pm$ 1.149 
& $\pm$ 1.265 & $\pm$ 1.156
\\
\hline
\ac{vanilla} & Opt &
33.490 & 22.120 & 13.523 & 4.058 & -5.876
\\
& 
& $\pm$ 4.490 & $\pm$ 4.808 & $\pm$	5.334 & $\pm$ 4.543 & $\pm$	4.421
\\
\hline
\acl{hybrid} & Full &
-1.417 & -11.397 & -21.383 & -31.545 & -41.395
\\
& 
& $\pm$ 1.244 & $\pm$ 1.248 & $\pm$	1.148 
& $\pm$	1.252 & $\pm$ 1.200
\\
\hline
\rowcolor{SeaGreen}
\acl{rn} & Full &
-1.790 & -11.847 & -21.738 & -31.620 & -41.831
\\
\rowcolor{SeaGreen}
& & $\pm$ 1.242 & $\pm$	1.190 & $\pm$ 1.281
& $\pm$	1.107 & $\pm$ 1.289
\\
\hline
\ac{kf} & Partial & 
11.154 & 0.926 & -9.160 & -18.428 & -28.786
\\
& 
& $\pm$ 3.023 & $\pm$ 3.064 & $\pm$	3.651 & $\pm$ 3.253 & $\pm$	3.301
\\
\hline
\ac{ks} & Partial & 
5.502 & -4.825 & -15.062 & -24.198 & -34.592
\\
&
& $\pm$ 2.942 & $\pm$ 3.205 & $\pm$ 3.440
& $\pm$ 3.119 & $\pm$ 3.133
\\
\hline
\acl{hybrid} & Partial & 
-0.989 & -11.123 & -20.865 & -30.080 & -38.174
\\
& 
& $\pm$ 1.223 & $\pm$ 1.243 & $\pm$	1.587
& $\pm$ 1.354 & $\pm$ 2.624  
\\
\hline
\acl{rn} & Partial & 
-0.774 & -10.852 &	-21.104 & -29.667 &	-38.066
\\
& & $\pm$ 1.243 & $\pm$ 1.158  & $\pm$	1.216
& $\pm$ 1.215 & $\pm$ 1.136
\\
\hline
\rowcolor{SeaGreen}
\acl{rn} & $\hat{\gvec{H}}$ & -1.743 & -11.697 & -21.721 & -31.186 & -40.301
\\
\rowcolor{SeaGreen}
&
& $\pm$ 1.269 & $\pm$	1.241 & $\pm$	1.153 & $\pm$	1.220 & $\pm$	1.169
\\
\hline
\end{tabular}
\label{tbl:lin_obs_miss}
}
\end{center}
\end{table}
%
%
%
%
%
\newpage
{\bf State Evolution Mismatch:}
We next consider a similar, but more challenging use case. Here, the \emph{design} evolution model is either \emph{unknown}, or assumed to be $\gvec{F}_{0}=\gvec{I}$, while the \emph{true} evolution model is
\begin{equation}\label{eqn:lin_evol_miss}
\gvec{F}_{\alpha^\circ}=\gvec{R}_{x,y}\brackets{\alpha}\cdot\gvec{F}_{0}, 
\quad
\alpha^\circ=10.
\end{equation}
The empirical results reported in \tbref{tbl:lin_evol_miss} demonstrate that while \ac{ks} experienced a severe performance degradation, \acl{rn} is able to compensate for unknown model information, by pre-estimating the evolution model via \ac{ls}, and achieve the \acl{lb}. We can again clearly notice the performance superiority of our \acl{rn} over its \ac{dd} counterparts, both for full and unknown information.
 
%
%
\begin{table}
\vspace{0.3cm}
\caption{Linear \ac{ss} - Evolution mismatch: $\nu=-20\dB$ }
\vspace{-0.3cm}
\begin{center}
\scriptsize{
\begin{tabular}{|c|c|c|c|c|c|c|c| }
\hline
\rowcolor{lightgray}\multicolumn{2}{|c|}{$\gscal{r}^{2} \dB $} & $10$ & $0$ & $-10$ & $-20$ & $-30$ \\
\hline
\ac{kf} & Full &
3.450 & -6.594 & -16.562 & -26.601 & -36.565
\\
& 
& $\pm$ 1.846 & $\pm$ 1.883	& $\pm$ 1.876 & $\pm$ 1.907	& $\pm$ 1.831
\\
\hline
\rowcolor{CornflowerBlue}
\ac{ks} & Full &
-3.843 & -13.913 & -23.592 & -33.861 &	-43.593
\\
\rowcolor{CornflowerBlue}
& 
& $\pm$ 2.655 & $\pm$ 2.709 & $\pm$ 2.751 & $\pm$	2.753 & $\pm$	2.746
\\
\hline
\ac{vanilla} & Opt & 
40.915 & 30.714 & 20.796 & 12.593 &	2.411
\\
& 
& $\pm$ 5.033 & $\pm$ 5.317 & $\pm$	4.955 & $\pm$ 4.281 & $\pm$	4.069
\\
\hline
\acl{hybrid} & Full & 
-1.975 & -11.850 & -21.403 & -30.579 & -41.016
\\
& 
& $\pm$ 2.549 & $\pm$ 3.369	& $\pm$ 2.564 & $\pm$ 2.548 
& $\pm$ 2.682
\\
\hline
\rowcolor{SeaGreen}
\acl{rn} & Full &
-3.351 & -13.585 & -23.333 & -33.126 & -43.160
\\
\rowcolor{SeaGreen}
& 
& $\pm$ 2.699 & $\pm$	2.673 & $\pm$	2.671 & $\pm$	2.628 & $\pm$	2.649
\\
\hline
\ac{kf} & Partial & 
33.961 & 23.833 & 13.848 & 4.199 & -6.434
\\
& 
& $\pm$ 3.933 & $\pm$ 3.857 & $\pm$	3.683 & $\pm$ 3.850	& $\pm$ 3.812
\\
\hline
\ac{ks} & Partial & 
32.963 & 22.838 & 12.853 & 3.201 & -7.431
\\
& 
& $\pm$ 3.933 & $\pm$ 3.859 & $\pm$ 3.685 & $\pm$ 3.851 & $\pm$ 3.809
\\
\hline
\acl{hybrid} & Partial & 
12.150 & -1.152 & -5.042 & -10.950 & -26.154
\\
& 
& $\pm$ 4.215 & $\pm$ 6.240 & $\pm$	4.272 & $\pm$ 2.790
& $\pm$ 3.968
\\
\hline
\acl{rn} & Partial & 
10.553 & -2.011 & -10.689 & -21.683 & -31.887
\\
& 
& $\pm$ 3.151 & $\pm$ 1.945 & $\pm$ 1.934 & $\pm$ 1.643 & $\pm$ 1.244
\\
\hline
\rowcolor{SeaGreen}
\acl{rn} & $\hat{\gvec{F}}$ &
-3.433 & -12.945 & -23.013 & -32.932 & -41.864
\\
\rowcolor{SeaGreen}
& 
& $\pm$ 2.633 & $\pm$	2.682 & $\pm$ 2.798 & $\pm$	2.471 & $\pm$ 2.657
\\
\hline
\end{tabular}
}
\label{tbl:lin_evol_miss}
\end{center}
\end{table}
%
%
%
\subsection{Kinematic Linear Differential Equations}\label{emp:lin_sde}
As a concluding experiment in a setting of linear \ac{ss} models, we consider smoothing in dynamics obtained from a \ac{sde} with a model mismatch. The state here represents a moving object obeying the \ac{ca} model~\cite{bar2004estimation} for one-dimensional kinematics. Here, $\gvec{x}_t=\brackets{p_t, v_t, a_t}^\top\in\greal^3$, where $p_t$, $v_t$, and $a_t$ are the position, velocity, and acceleration, respectively, at time $t$. We observe noisy position measurements sampled at time intervals $\Delta{t}=10^{-2}$, yielding a linear Gaussian \ac{ss} model with  $\gvec{H}=\brackets{1, 0, 0}$ and 
\begin{equation*}
\gvec{F}\!=\!
\begin{pmatrix}
1 & \Delta\tau & \frac{1}{2}\Delta\tau^2 \\
0 & 1 & \Delta\tau \\
0 & 0 & 1
\end{pmatrix};
\gvec{Q}\!=\!
\gscal{q}^2\cdot\begin{pmatrix}
\frac{1}{20}\Delta\tau^5 & 
\frac{1}{8}\Delta\tau^4  & 
\frac{1}{6}\Delta\tau^3 \\
\frac{1}{8}\Delta\tau^4 & \frac{1}{3}
\Delta\tau^3 &
\frac{1}{2}\Delta\tau^2 \\
\frac{1}{6}\Delta\tau^3 &
\frac{1}{2}\Delta\tau^2 &
\Delta\tau
\end{pmatrix}. 
\end{equation*}

While for the synthetic linear models considered in the previous subsections we used \acl{rn} with a single forward-backward pass, here we evaluate it with $K=2$ unfolded pass, comparing it to both the \ac{ks} and to \acl{hybrid} when recovering the entire state vector, as well as when recovering only the position (which is often the case in positioning applications). For the latter, we also consider the case where the smoothers assume a more simplified \ac{cv} model~\cite{bar2004estimation} state evolution for state evolution. The \ac{cv} model captures in its state vector the position and velocity (without the acceleration), and is a popular model for kinematics due to its simplicity. Yet, for the current setting, it induces an inherent model mismatch. The results are reported in~\tbref{tbl:lin_sde}. 

For the \acl{hybrid}  smoother of~\cite{satorras2019combining}, which is \ac{dd} yet also requires knowledge of the \ac{ss} model, we optimized $\gvec{Q}$ via grid search to achieve the best performance, as it was shown to be unstable when substituting the true $\gvec{Q}$. In~\tbref{tbl:lin_sde}, we observe that \acl{rn} comes within \textcolor{NewColor}{a} minor gap of the \ac{ks} in estimating both the full state as well as only the positions when it is known that the state obeys the \ac{ca} model; When it is postulated that the state obeys the \ac{cv} model, \acl{rn} outperforms all benchmarks.

\begin{table}
\vspace{0.3cm}
\caption{Linear kinematic \ac{ss} model}
\vspace{-0.3cm}
\begin{center}
\scriptsize{
\begin{tabular}{|c|c|c|>{\columncolor{CornflowerBlue}}c|c|>{\columncolor{SeaGreen}}c|}
\hline
\rowcolor{lightgray}
Model & Error & \ac{kf} & \ac{ks} & \acl{hybrid} & \acl{rn}-2 \\
\hline
\ac{ca} & Full State &
-7.631 & -8.791 & 14.351 & -8.432
\\
&
& $\pm$ 2.891 & $\pm$ 3.054	& $\pm$ 2.011 & $\pm$ 2.974
\\
\hline
\ac{ca} & Position &
-22.074 & -23.221 & -11.456 & -22.241
\\
&
& $\pm$ 3.694 & $\pm$ 4.081	& $\pm$ 2.037 & $\pm$ 3.676	
\\
\hline
\ac{cv} & Position &
-7.657 & -14.752 & -10.732 & -15.900
\\
&
& $\pm$ 3.145 & $\pm$ 3.308	& $\pm$ 1.661 & $\pm$ 2.542
\\
\hline
\end{tabular}
}
\label{tbl:lin_sde}
\end{center}
\end{table}

\color{NewColor}
In the study reported in Table~\ref{tbl:lin_sde}, the state trajectory was simulated from a kinematic \ac{ca} model, which is a linear Gaussian state evolution model. We next show that the improved performance of \acl{rn} is preserved also when tracking states corresponding to real-world vehicular trajectories, that are only approximated by linear Gaussian models. To that aim, we use the city recordings from the KITTI data set~\cite{geiger2013vision}, where each sample represents the position of a vehicle in three-dimensional space, with 16 training trajectories, 2 validation trajectories, and 6 testing trajectories, all sampled at intervals of $\Delta t = 10^{-2}$. The measurements are noisy observations of the position corrupted by Gaussian noise with covariance $\gmat{R}=\gmat{I}_3$

In Table~\ref{tbl:KITTI} we compare the \ac{mse} achieved by \acl{rn} with $K=1$ to that of the \ac{mb} \ac{kf} and \ac{ks}, where all smoothers assume a \ac{cv} model on the underlying state trajectory. We observe in Table~\ref{tbl:KITTI} that \acl{rn} outperforms the \ac{mb} benchmarks, as it learns from data to compensate for the inherent mismatch in modeling the underlying real-world vehicular trajectory as obeying a \ac{cv} model.

\begin{table}
\vspace{0.3cm}
\caption{KITTI  kinematic \ac{ss} model}
\vspace{-0.3cm}
\begin{center}
\scriptsize{
\begin{tabular}{|c|c|>{\columncolor{CornflowerBlue}}c|>{\columncolor{SeaGreen}}c|}
\hline
\rowcolor{lightgray}
Model & \ac{kf} & \ac{ks}  & \acl{rn}-1 \\
\hline
\ac{cv} & 
-21.395 & -25.158 &-26.566
\\
&
 $\pm$ 0.486 & $\pm$ 0.633 & $\pm$ 0.422
\\
\hline
\end{tabular}
}
\label{tbl:KITTI}
\end{center}
\end{table}

\color{black}

\vspace{0.3cm}
\subsection{Nonlinear Lorenz Attractor}\label{emp:Lorenz}
We proceed to evaluate \acl{rn} in a \acl{nl} \ac{ss} model following the \acl{la}, which is a three-dimensional chaotic solution to the Lorenz system of ordinary differential equations.
This synthetically generated chaotic system exemplifies dynamics formulated with \acp{sde}, that demonstrates the task of smoothing a highly \acl{nl} trajectory and a \acl{rw} practical challenge of handling mismatches due to sampling a {\acl{ct}} signal into \acl{dt}~\cite{gilpin2021chaos}. As the dynamics are \acl{nl}, here we use \acl{rn} with both $K=1$ and $K=2$ forward-backward passes, denoted \acl{rn}-1 and \acl{rn}-2, respectively. 

The \acl{la} models the movement of a particle in 3D space, i.e., $m=3$, which, when sampled at  interval $\Delta \tau$, obeys a state evolution model with $\gvec{f}\brackets{\gvec{x}_t}=
\gvec{F}\brackets{{\gvec{x}_t}}\cdot\gvec{x}_t$, where 
\begin{equation}\label{eqn:Taylor}
\gvec{F}\brackets{{\gvec{x}_\tau}}=
\gvec{I} + \sum_{j=1}^{J} \frac{\brackets{\gvec{A}\brackets{{\gvec{x}_\tau}}\cdot\Delta \tau}^j}{j!},
\end{equation}
with $J$ denoting the order of the Taylor series approximation used to obtain the model (where we use $J=5$ when generating the data), and 
\begin{equation*}
\gvec{A}\brackets{{\gvec{x}_\tau}}=
\begin{pmatrix}
-10 & 10 & 0\\
28 & -1 & -\gscal{x}_{1, \tau}\\
0 & \gscal{x}_{1, \tau} & -\frac{8}{3}
\end{pmatrix}.
\end{equation*}
We first evaluate \acl{rn} under noisy rotated state observations, with and without observation model mismatch as well as with sampling mismatches, after which we evaluate it with \acl{nl} observations. 

%
\begin{table}
\vspace{-0.3cm}
\caption{\acl{la} - Observation Mismatch: }
\vspace{-0.3cm}
\begin{center}
\scriptsize{
\begin{tabular}{|c|c|c|c|c|c|c|c| }
\hline
\rowcolor{lightgray}
\multicolumn{2}{|c|}{$\gscal{r}^{2} \dB $} & $10$ & $0$ & $-10$ & $-20$ & $-30$ \\
\hline
\multicolumn{2}{|c|}{Noise} & 
10.017 & 0.005 & -10.011 & -19.942 & -29.986
\\
\multicolumn{2}{|c|}{} 
& $\pm$ 0.334  & $\pm$	0.376  & $\pm$ 0.368  & $\pm$ 0.347	 & $\pm$ 0.354
\\
\hline
\acs{ekf} & Full & 
-0.299 & -10.533 & -20.493 & -30.348 & -40.483
\\
& & 
$\pm$ 1.084 & $\pm$ 1.016	& $\pm$ 0.969 & $\pm$ 1.016 & $\pm$	0.992
\\
\hline
\rowcolor{CornflowerBlue}
\ac{eks} & Full & 
-3.892 & -13.752 & -23.868 & -33.743 & -43.755
\\
\rowcolor{CornflowerBlue}
&  
& $\pm$ 0.996 & $\pm$ 1.161 & $\pm$	1.025 & $\pm$ 1.013 & $\pm$ 1.145
\\
\hline
\acl{hybrid} & Full &
-2.263 & -12.398 & -22.413 & -31.040 & -42.368
\\
& 
& $\pm$ 1.113 & $\pm$	1.182 & $\pm$	1.076 & $\pm$	1.046 & $\pm$	1.256
\\
\hline
\rowcolor{SeaGreen}
\acl{rn}-2 & Full & 
-3.138 & -13.330 & -23.304 & -33.311 & -43.235
\\
\rowcolor{SeaGreen}
& 
& $\pm$ 0.983 & $\pm$ 1.195 & $\pm$	1.036 & $\pm$ 0.999 & $\pm$ 1.112
\\
\hline
\acs{ekf} & Partial & 
-0.258 & -9.747 & -15.945 &	-17.549 & -17.752
\\
& 
& $\pm$ 1.073 & $\pm$ 0.988 & $\pm$ 0.769 & $\pm$ 0.325 & $\pm$	0.109
\\
\hline
\ac{eks} & Partial & 
-3.824 & -12.932 & -18.957 & -20.363 & -20.563
\\
&
& $\pm$ 0.976 & $\pm$ 1.122 & $\pm$	0.853 & $\pm$ 0.366	& $\pm$ 0.126
\\
\hline
\acl{hybrid} & Partial & 
-1.921 & -11.959 & -18.724 & -23.076 & -23.351
\\
& 
& $\pm$ 0.962 & $\pm$ 1.308	& $\pm$ 0.871 & $\pm$ 0.743	& $\pm$ 0.472 \\
\hline
\acl{rn}-2 & Partial & 
-3.010 & -13.290 & -22.620 & -31.789 & -41.874
\\
&  
& $\pm$ 1.067 & $\pm$ 1.175 & $\pm$ 1.077 & $\pm$ 1.207 & $\pm$	1.216
\\
\hline
\rowcolor{SeaGreen}
\acl{rn}-2 & $\hat{\gvec{H}}$ & 
-3.127 & -13.315 & -23.158 & -32.581 & -42.928
\\
\rowcolor{SeaGreen}
&
& $\pm$ 1.057 & $\pm$ 1.194 & $\pm$ 1.043 & $\pm$ 1.119 & $\pm$ 1.145
\\
\hline
\end{tabular}
\label{tbl:Lorenz_obs_rot}
}
\end{center}
\end{table}
%
%
\begin{figure}
\vspace{-0.35cm}
\includegraphics[width=1\columnwidth]{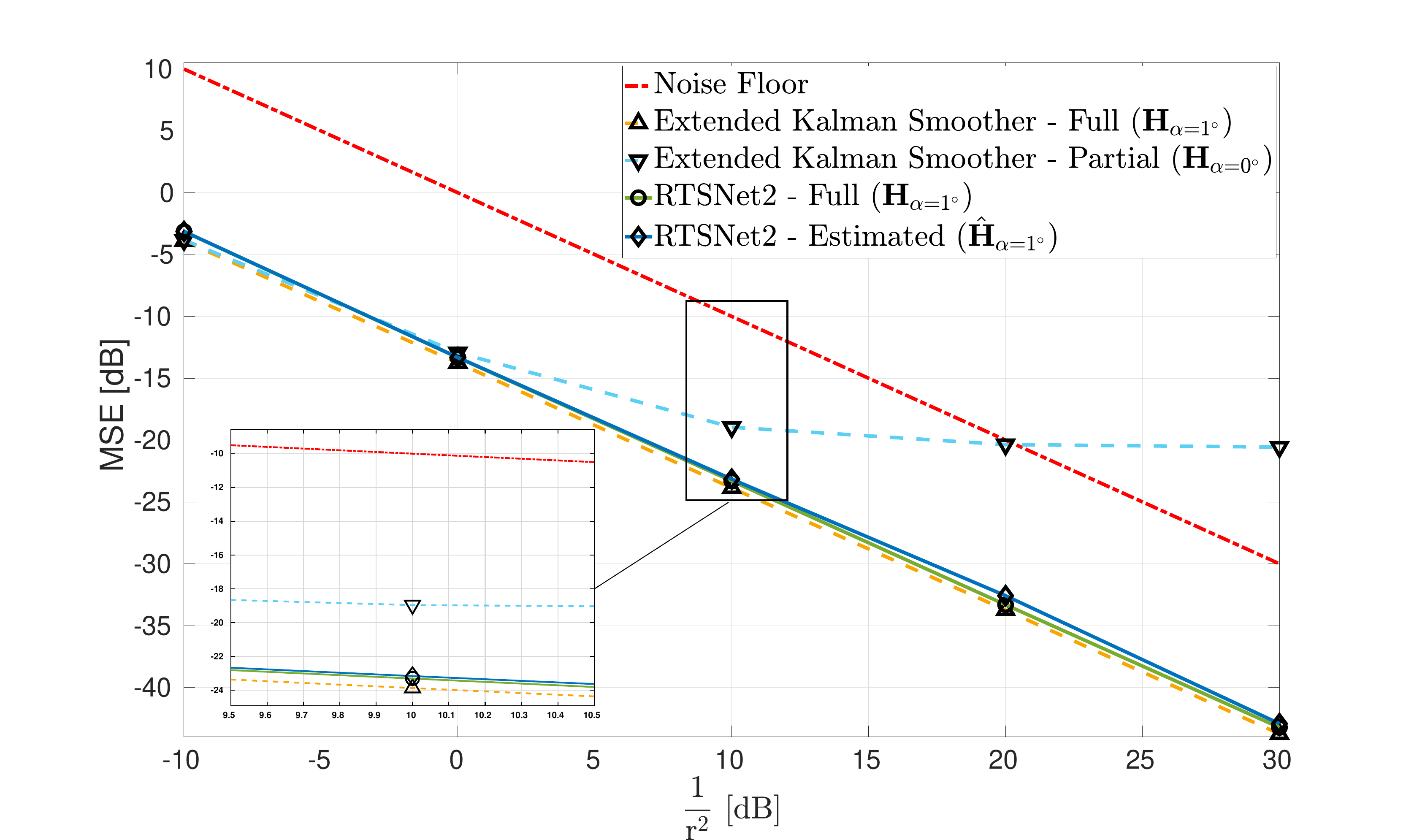}
\caption{$T=2000$, $\nu=-20\dB$, $\gobs=\gvec{I}$.}
\label{fig:sim_Lorenz_Full_H_rot}
\end{figure}
%
%

\smallskip
{\bf Rotated State Observations:}
Here, we consider the \acl{dt} state evolution with $\Delta{\tau}=0.02$. The observations model 
 $\gobs$ is set to a rotation matrix with $\alpha=1^\circ$, whereas $T=100$ and $\nu=-20\dB$. 
As in Subsection~\ref{ssec:partialLinear}, we consider the cases where the smoothers are aware of the rotation (Full information); when the assumed state evolution is the identity matrix instead of the slightly rotated one (Partial information), as well as when it is estimated from the data set via \ac{ls} ($\hat{\gvec{H}}$). 

The results reported in \tbref{tbl:Lorenz_obs_rot} and in \figref{fig:sim_Lorenz_Full_H_rot} demonstrate that although \acl{rn} does not have access to the true statistics of the noise, for the case of \emph{full} observation model information, it still achieves the \ac{mse} \acl{lb}. It is also observed that a mismatched state observation model obtained from a seemingly minor rotation causes severe performance degradation for the \ac{ks}, which is sensitive to model uncertainty, while \acl{rn} is able to learn from data to overcome such mismatches. Finally, empirical observations reveal that \acl{rn} consistently surpasses the \ac{dd} benchmark presented in~\cite{satorras2019combining}. Furthermore, the utilization of its unfolding mechanism with $K=2$ forward-backward passes leads to a notable enhancement in performance compared to just a single pass.
%
%

\begin{figure*}
\vspace{-0.2cm}
\begin{center}
%
%
\begin{subfigure}[pt]{0.99\columnwidth}
\includegraphics[width=1\columnwidth]{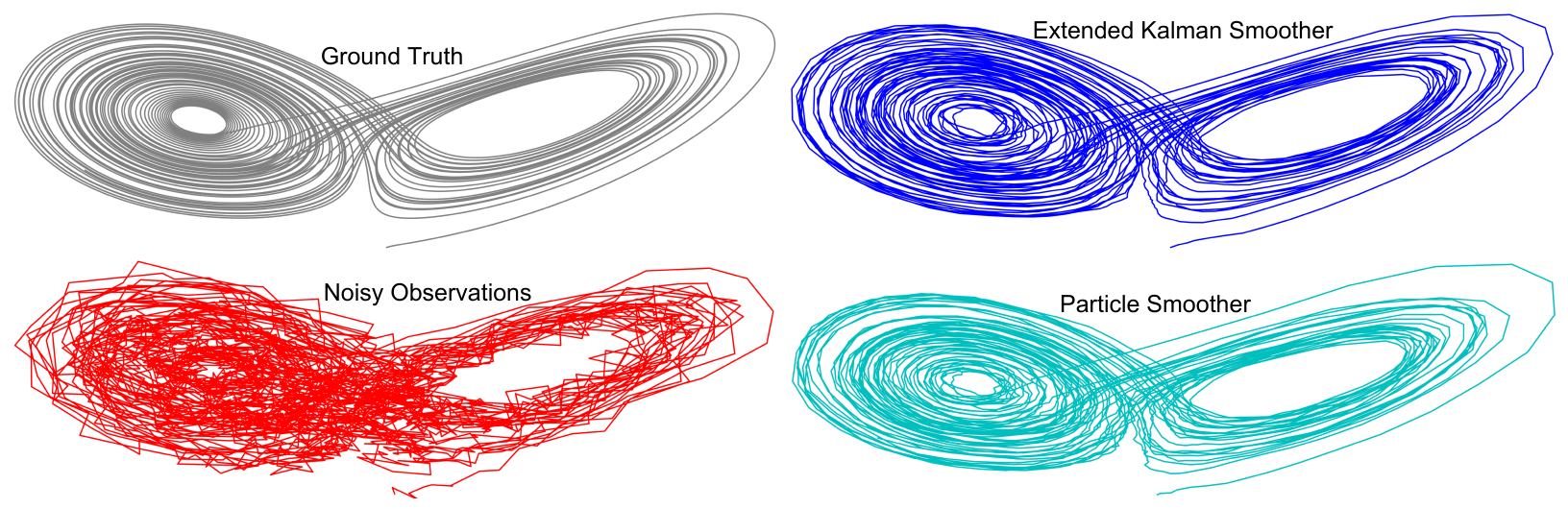}
\end{subfigure}
%
%
\begin{subfigure}[pt]{0.99\columnwidth}
\includegraphics[width=1\columnwidth]{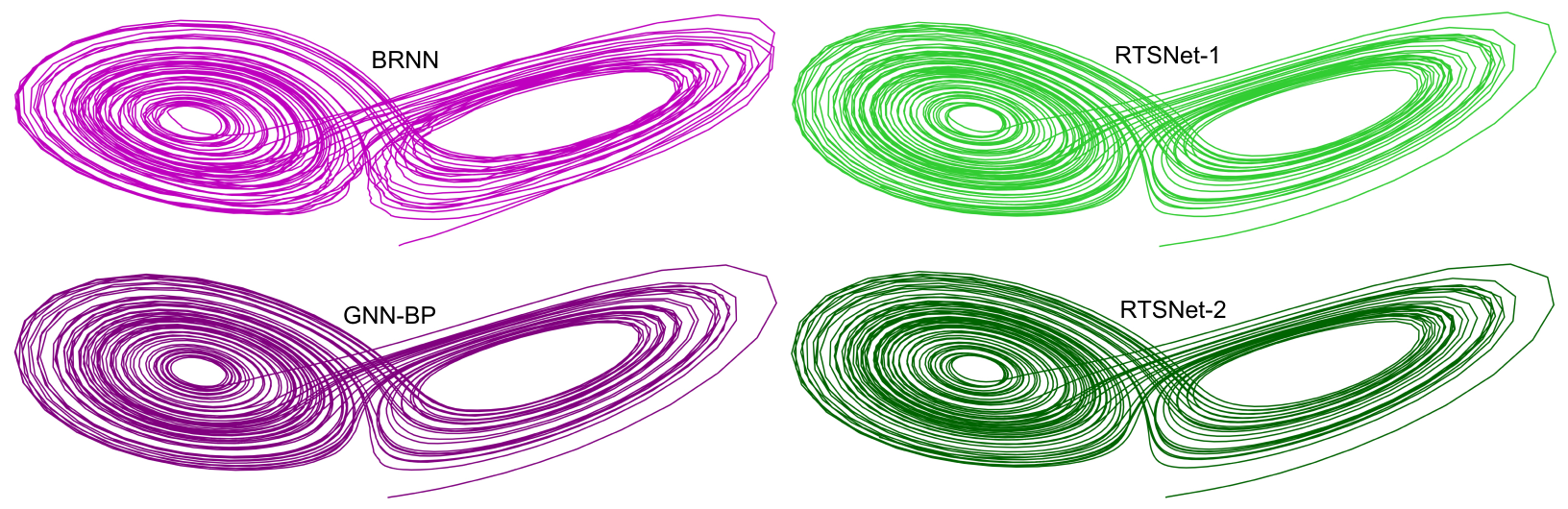}
\end{subfigure}
\caption{\acl{la} with sampling mismatch (decimation), $T=3000$.}
\label{fig:decimation}
\end{center}
\figSpace
\vspace{-0.2cm}
\end{figure*}

{\bf Sampling Mismatch:}
Here, we demonstrate a practical case where a physical process evolves in \acl{ct}, but the smoother only has access to noisy observations in \acl{dt}, which then results in an inherent mismatch in the \ac{ss} model. We generate data from an approximate \acl{ct} noise-less state evolution $\gvec{F}\brackets{{\gvec{x}_\tau}}$, with high resolution time interval ${\Delta\tau=10^{-5}}$. We then sub-sampled the process by a ratio of $\frac{1}{2000}$ and get a decimated process with $\Delta{t}=0.02$. Finally, we generated noisy observations of the true state, by using an identity observation matrix $\gobs=\gvec{I}$ and non-correlated observation noise with $\gscal{r}^2=0\dB$. See an example in \figref{fig:decimation}: \acl{gt}, and noisy observations, respectively.

The \ac{mse} values for smoothing sequences with length $T=3000$, reported in \tbref{tbl:decimation}, demonstrate that \acl{rn} overcomes the mismatch induced by representing a \acl{ct} \ac{ss} model in \acl{dt}, achieving a substantial processing gain over \textcolor{NewColor}{its} \ac{mb} and \ac{dd} counterparts due to its learning capabilities. In \figref{fig:decimation}, we visualize how this gain is translated into clearly improved smoothing of a single trajectory. 
%
%
\begin{table*}
\begin{center}
\caption{\ac{mse} $\dB$ - \acl{la} with sampling mismatch.}
\scriptsize{
\begin{tabular}{|c|c|c|c|c|c|c|c|c|c|c|}
\hline
\rowcolor{lightgray}
Noise & \ac{ekf} & \acs{pf} & \acl{kn} &
\ac{eks} & \ac{ps} & \ac{vanilla}  & \acl{hybrid} & \acl{rn}-1 & \acl{rn}-2 
\\
\hline
-0.024 & -6.316  & -5.333 & -11.106 & 
-10.075 & -7.222 & -2.342 & -16.479 & -15.436 &  \cellcolor{SeaGreen}-16.803
\\
$\pm$ 0.049 & $\pm$ 0.135 & $\pm$ 0.136 & $\pm$ 0.224 & 
$\pm$ 0.191 & $\pm$ 0.202 & $\pm$ 0.092 & $\pm$ 0.352 & $\pm$ 0.329 &  \cellcolor{SeaGreen} $\pm$ 0.301 
\\
\hline
\end{tabular}
\label{tbl:decimation}
}
\end{center}
\vspace{-0.2cm}
\end{table*} 
%
%
%

%

{\bf Noisy Nonlinear Observations:}
Finally, we consider the case of the \acl{dt} \acl{la},  with \acl{nl} observations, which take the form of a transformation from a cartesian coordinate system to spherical coordinates. In such settings, the observations function is given by

\begin{equation*}
\gvec{h}
\brackets{\sbrackets{\gscal{x}, \gscal{y}, \gscal{z}}^\top} \triangleq 
\sbrackets{
\begin{array}{c}
\sqrt{\gscal{x}^2 + \gscal{y}^2 + \gscal{z}^2}  \\
\tan ^{-1} \brackets{\frac{\gscal{y}}{\gscal{x}}} \\
\cos ^{-1}\brackets{{\frac{\gscal{z}}{\sqrt {\gscal{x}^2 + \gscal{y}^2 + \gscal{z}^2}}}}    
\end{array}}.
\end{equation*}
We further set $T=20$ and $\nu=0\dB$. 

\begin{figure}
\vspace{-0.325cm}
\includegraphics[width=1\columnwidth]{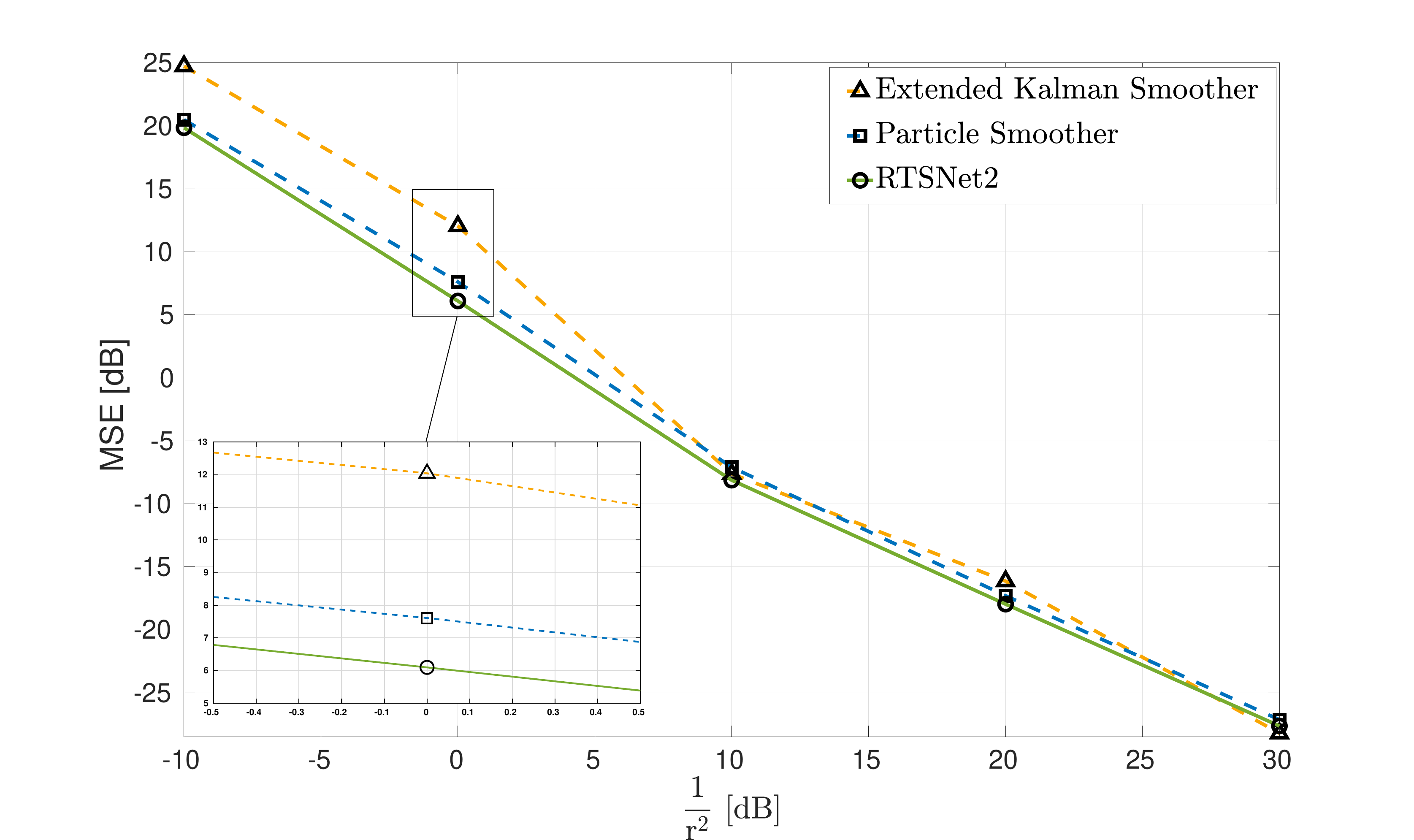}
\caption{$T=20$, $\nu=0\dB$, $\gobs$ \acl{nl}.}
\label{fig:sim_Lorenz_full_NL}
\end{figure}

The \ac{mse} achieved by \acl{rn} with $K=2$ forward-backward passes is compared with that of the \ac{ks}, \ac{ps} and \acl{rn} with $K=1$, reported in \tbref{tbl:Lorenz_obs_NL} and depicted in \figref{fig:sim_Lorenz_full_NL}. It is clearly observed here that in such \acl{nl} setups, \acl{rn} outperforms its \ac{mb} counterparts which operate with full knowledge of the underlying \ac{ss} model, indicating \textcolor{NewColor}{t}he ability of its \ac{dnn} augmentation and unfolded architecture to improve performance in the presence of nonlinearities.

%
%
\begin{table}
\vspace{-0.2cm}
\caption{\acl{la} with \acl{nl} observations}


\vspace{-0.3cm}
\begin{center}
\scriptsize{
\begin{tabular}{|c|c|c|c|c|c|c|c|}
\hline
\rowcolor{lightgray}
\multicolumn{2}{|c|}{$\gscal{r}^{2} \dB $} & $10$ & $0$ & $-10$ & $-20$ & $-30$ \\
\hline
\ac{ekf} & Full & 
24.693 & 12.197 & -6.343 & -15.574 & -26.418
\\
& 
& $\pm$ 4.147 & $\pm$ 8.061 & $\pm$	1.961 & $\pm$ 3.451 & $\pm$	1.743
\\
\hline
\rowcolor{CornflowerBlue}
\ac{eks} & Full & 
24.739 & 12.045 & -7.613 & -16.134 & -28.211
\\
\rowcolor{CornflowerBlue}
& 
& $\pm$ 4.313 & $\pm$ 8.260 & $\pm$	2.474 & $\pm$ 5.157 & $\pm$ 1.548
\\
\hline
\ac{ps} & Full & 
20.490 & 7.612 & -7.093 & -17.293 &	-27.138
\\
& 
& $\pm$ 6.187 & $\pm$ 10.071 & $\pm$ 1.822 & $\pm$ 1.704 & $\pm$	1.743
\\
\hline
\acl{rn}-1 & Full & 
21.094 & 10.804 & -8.074 & -17.941 & -27.476
\\
& 
& $\pm$ 2.901 & $\pm$ 8.999 & $\pm$	1.500 & $\pm$ 1.712 & $\pm$	1.553
\\
\hline
\rowcolor{SeaGreen}
\acl{rn}-2 & Full & 
19.849 & 6.100 & -8.122 & -17.960 & -27.630
\\
\rowcolor{SeaGreen}
& 
& $\pm$ 4.183 & $\pm$ 6.614 & $\pm$	1.521 & $\pm$ 1.676 & $\pm$	1.558
\\
\hline
%
\end{tabular}
\label{tbl:Lorenz_obs_NL}
}
\end{center}
\vspace{-0.325cm}
\end{table}
%
%
%
\color{NewColor}
\subsection{Nonlinear Van Der Pol Oscillator}\label{ssec:vanderpol}
The study reported in Subsection~\ref{emp:Lorenz} shows the ability of \acl{rn} to successfully cope with harsh nonlinearities in the \ac{ss} model. To further demonstrate this property of \acl{rn}, we next evaluate it in tracking the Van Der Pol Oscillator~\cite[Sec. 4.1]{kandepu2008applying}, where the state is a two-dimensional vector governed by the following \acl{nl} state-evolution model
\begin{equation}
\label{eqn:vdp}
\gvec{f}\brackets{\gvec{x}_{\tau}} = 
\brackets{
\begin{array}{c}
\gscal{x}_{1,\tau} +\gscal{x}_{2,\tau}\cdot \Delta \tau \\
\gscal{x}_{2,\tau} + 
\brackets{2\brackets{1-\gscal{x}_{1,\tau}^2}\cdot \gscal{x}_{2,\tau} -\gscal{x}_{1,\tau}}\Delta \tau 
\end{array}},
\end{equation}
with $\Delta \tau = 0.1$
Tracking is done based on noisy observations of the first state element, i.e., 
$\gmat{H} = (1, 0)$, with $\gmat{Q}=0.01\cdot \gmat{I}_2$ and $\gmat{R}=1$. The initial state is fixed to $\gvec{x}_0=(0, -5)^\top$, and the trajectory length is $T=40$.

In addition to comparing \acl{rn} to the \ac{mb} \ac{ekf} and \ac{eks}, here we also compare it to optimization-based smoothers that are derived from the \acs{map} formulation,  and particularly the \ac{mb} Gauss-Newton method of~\cite[Sec. 3]{aravkin2014optimization}. This optimization-based smoother is typically capable of tracking in \acl{nl} \ac{ss} models, while coinciding with the \ac{mse} optimal \ac{eks} for linear Gaussian cases. The resulting \ac{mse} values are reported in Table~\ref{tbl:Oscillator}. There, it is observed that the \acl{nl} state evolution model in \eqref{eqn:vdp} limits the performance of the \ac{ekf} and the \ac{eks}, which are outperformed by the Gauss-Newton method of~\cite{aravkin2014optimization}. Still, \acl{rn} is shown in Table~\ref{tbl:Oscillator} to outperform all these \ac{mb} smoothers, which operate with full knowledge of the underlying \ac{ss} model and the noise distribution.

\begin{table}
\vspace{0.3cm}
\caption{Van der Pol oscillator}
\vspace{-0.3cm}
\begin{center}
\scriptsize{
\begin{tabular}{|c|>{\columncolor{CornflowerBlue}}c|c|>{\columncolor{SeaGreen}}c|}
\hline
\rowcolor{lightgray}
\ac{ekf} & \ac{eks} & Gauss-Newton  & \acl{rn}-1 \\
\hline
12.711 & 
3.164 & -4.94 &-7.689
\\
$\pm$ 5.951&
 $\pm$ 4.135 & $\pm$ 2.45 & $\pm$ 3.102
\\
\hline
\end{tabular}
}
\label{tbl:Oscillator}
\end{center}
\end{table}

\color{black}

\subsection{Complexity Analysis}\label{ssec:complexity}
So far, we have demonstrated that \acl{rn} delivers superior \ac{mse} performance, surpassing both its \ac{dd} and \ac{mb} counterparts, especially when they operate with partial information or under \acl{nl} dynamics. We conclude our empirical study by highlighting that the advantages of \acl{rn} don't come with added computational complexity during inference, dependency on large datasets, or an increase in \ac{dnn} size

In \tbref{tbl:run_time}, we detail the average inference time for all filters (without parallelism) using the \acl{la} \acl{se} task as a benchmark. The stopwatch timings, captured on \textit{Google Colab} equipped with a CPU: Intel(R) Xeon(R) CPU @ 2.20GHz and GPU: Tesla P100-PCIE-16GB, reveal that \acl{rn} is highly competitive when compared with classical methods and even outperforms \acl{hybrid}. This superiority is primarily attributed to its efficient \acl{nn} computations. Furthermore, unlike the \ac{mb} filters, \acl{rn} bypasses the need for linearization and matrix inversions at each time step.

\begin{figure}
\includegraphics[width=1\columnwidth]{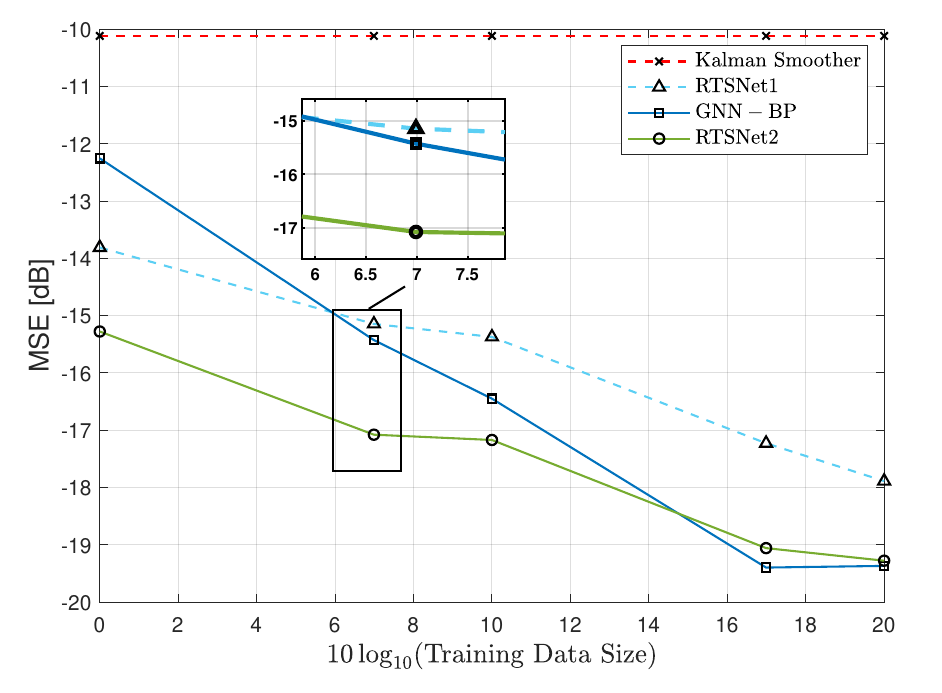}
\caption{\ac{mse} versus $10\log_{10} (N)$, \acl{nl}.}
\label{fig:Datasize}
\end{figure}

Subsequently, we delve into the volume of training trajectories required to effectively train \acl{rn}. Focusing again on the \acl{la} setup, we measure the average \ac{mse} against varying data sizes. Specifically, each \ac{dd} smoothing algorithm is trained with a varying number of trajectories (denoted as $N$) spanning a length of $T=3000$ pairs with an observation noise of $\gscal{r}^2= 0 \dB$. As depicted in \figref{fig:Datasize}, which showcases test \ac{mse} versus $10\log_{10}\brackets{N}$, \acl{rn} is successfully trained even with a single trajectory. Notably, its performance is approached by \ac{dd} \acl{hybrid} only when the number of trajectories exceeds $50$.

In conclusion, a side-by-side comparison of \ac{dnn} size, represented by the number of trainable parameters, between \acl{rn} and \acl{hybrid} for select use cases reveals some insightful findings. Table {\tbref{tbl:network_size}} lists the number of parameters, illustrating the compactness of \acl{rn}, thereby implying that it is easier to train. Significantly, \acl{rn} consistently outperforms the \ac{dd} \acl{hybrid} benchmark of~\cite{satorras2019combining} across tested scenarios. This is achieved with a simpler architecture that has fewer trainable parameters. This stands out especially considering that the design of~\cite{satorras2019combining} also emphasizes compactness and efficiency. Ultimately, The reduced parameterization of \acl{rn} leads to faster training and inference.

%
%
\begin{table*}
\begin{center}
\caption{Inference Time $\sbrackets{\sec}$ - \acl{la}.}
\scriptsize{
\begin{tabular}{|c|l|c|c|c|c|c|c|c|c|c|}
\hline
\rowcolor{lightgray}
$\textit{Use Case}$ & Trajectory Length &
\ac{kf} & \acs{pf} & \acl{kn} &
\ac{ks} & \ac{ps} & \acl{hybrid} & \acl{rn}-1 & \acl{rn}-2 \\
\hline
Nonlinear Observations & $T=20$
& 0.0501 & NA & NA 
& 0.0946 & 5.0175 & NA & 0.0605 & 0.1178 
 \\
\hline
Linear Observations & $T=100$
& 0.2194 & NA & NA 
& 0.4344 & 24.4158 & 1.2513 &  0.2950 & NA
\\
\hline
Decimation $\brackets{K=2}$ & $T=3000$
& 4.3583 & 45.4791 & 4.9226 
& 6.5164 & 452.8513 &  25.4527 & 7.3587 &  14.6174
\\
\hline
Decimation $\brackets{K=5}$ & $T=3000$
& 6.2641 & 71.6549 & NA 
& 10.3243 & 723.9320 &  NA & NA & NA
\\
\hline
\end{tabular}
\label{tbl:run_time}
}
\end{center}
\end{table*} 
%
%
\begin{table*}
\vspace{-0.3cm}
\begin{center}
\caption{Network Size - Number of Trainable Parameters}
\scriptsize{
\begin{tabular}{|c|c|c|c|}
\hline
\rowcolor{lightgray}
$\textit{Use Case}$ &
Linear - $2\times2$ & Linear - $5\times5$ & Lorentz - Decimation
\\
\hline
\acl{rn}
& $7,370$ & $28,285$ & 
$33,270~(\#1)~,~66,540~(\#2) $
 \\
\hline
\acl{hybrid} 
& $40,947$ & $41,814$ & $41,236$
\\
\hline
\end{tabular}
\label{tbl:network_size}
}
\end{center}
\end{table*} 
%
%
\section{Conclusion}\label{sec:Conclusion}
In this work, we introduced \acl{rn}, a hybrid fusion of \acl{dl} with the classic \ac{ks}. Our design identifies the \ac{ss}-model-dependent computations of the \ac{ks} and replaces them with dedicated \acp{rnn}. These \acp{rnn} operate on specific features that encapsulate the information necessary for their operation. Additionally, we unfold the algorithm to enable multiple trainable forward-backward passes. Our empirical studies reveal that \acl{rn} can perform offline state estimation similarly to the \ac{ks}, but with the added ability to learn to overcome model mismatches and nonlinearities. Notably, \acl{rn} employs a relatively compact \ac{rnn}, which can be trained with a modest-sized \acl{ds}, leading to reduced complexity.
%
%
%
\bibliographystyle{IEEEtran}
\bibliography{IEEEabrv,RTSNet}
%
%
\begin{IEEEbiography} [{\includegraphics[width=1in,height=1.25in,clip,keepaspectratio]{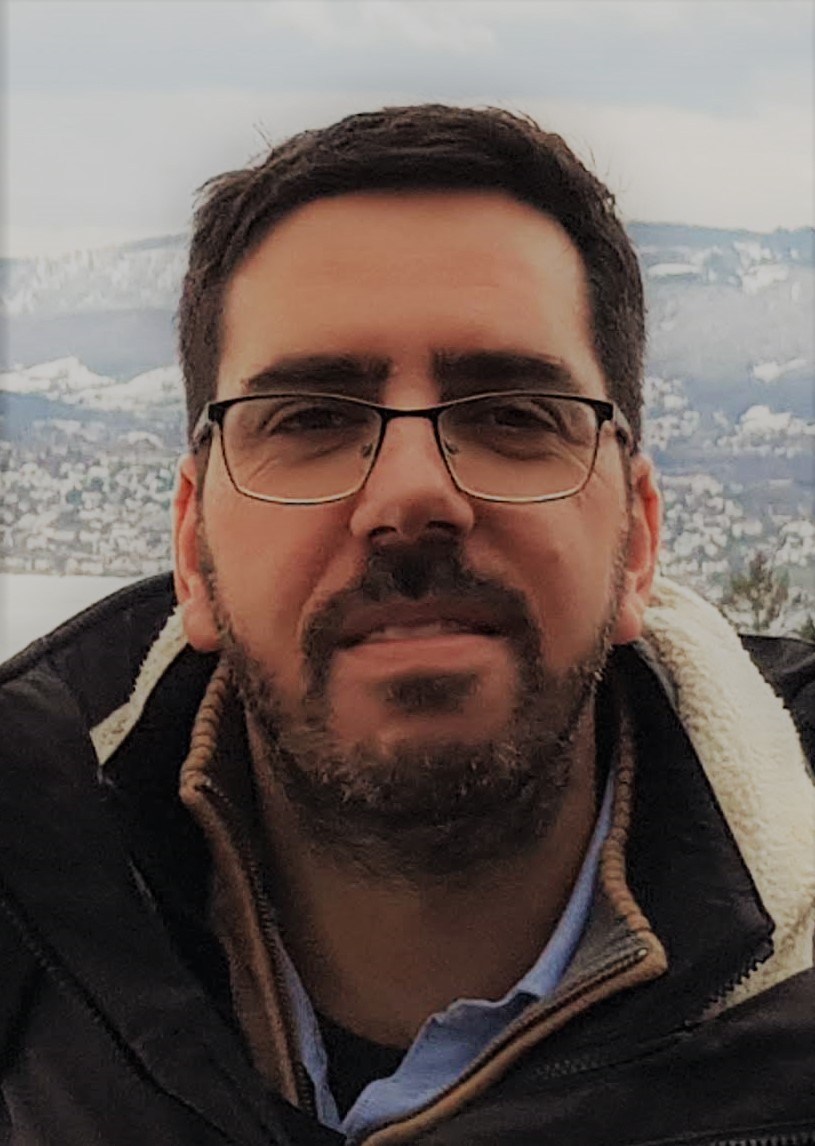}}] {Guy Revach} is a researcher with a proven industry track record as an innovator and system engineer. He received his B.Sc. (cum laude) and M.Sc. degrees in 2008 and 2017, respectively, from the Andrew and Erna Viterbi Department of Electrical \& Computer Engineering at the Technion – Israel Institute of Technology in Haifa. He completed his master’s thesis under the supervision of Prof. Nahum Shimkin on planning for cooperative multi-agents. Since 2019, he has been a Ph.D. candidate at the Institute for Signal and Information Processing (ISI) at ETH Zürich, Switzerland, supervised by Prof. Dr. Hans-Andrea Loeliger. His main research focus is on the intersection of machine learning with signal processing, specifically combining sound theoretical principles from classical signal processing with state-of-the-art machine learning algorithms for tracking and detection problems. Before joining ETH Zürich, he worked in the Israeli wireless communication industry for over 10 years, initially as a real-time embedded software engineer and later as a software manager. He was the main innovator behind state-of-the-art, software-defined radio (SDR) for wireless communication, which was game-changing and groundbreaking in terms of size, weight, and power. As a system engineer, he defined major aspects of SDR requirements and architecture, including hardware, software, network, cyber defense, signal processing, data analysis, and control algorithms.
\end{IEEEbiography} 

\begin{IEEEbiography} [{\includegraphics[width=1in,height=1.25in,clip,keepaspectratio]{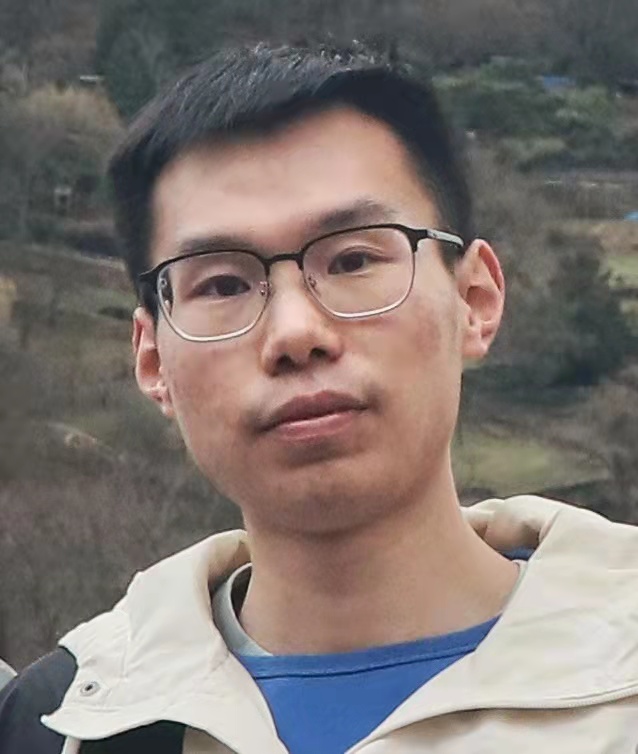}}] {Xiaoyong Ni} received a B.S. degree in Communication Engineering in 2020 from the University of Electronic Science and Technology of China (UESTC) in Chengdu, China. He received an M.S. degree from the Department of Electrical Engineering and Information Technology at ETH Zürich. His current research interests include signal processing, machine learning, and wireless communication.
\end{IEEEbiography} 

\begin{IEEEbiography}[{\includegraphics[width=1in,height=1.25in,clip,keepaspectratio]{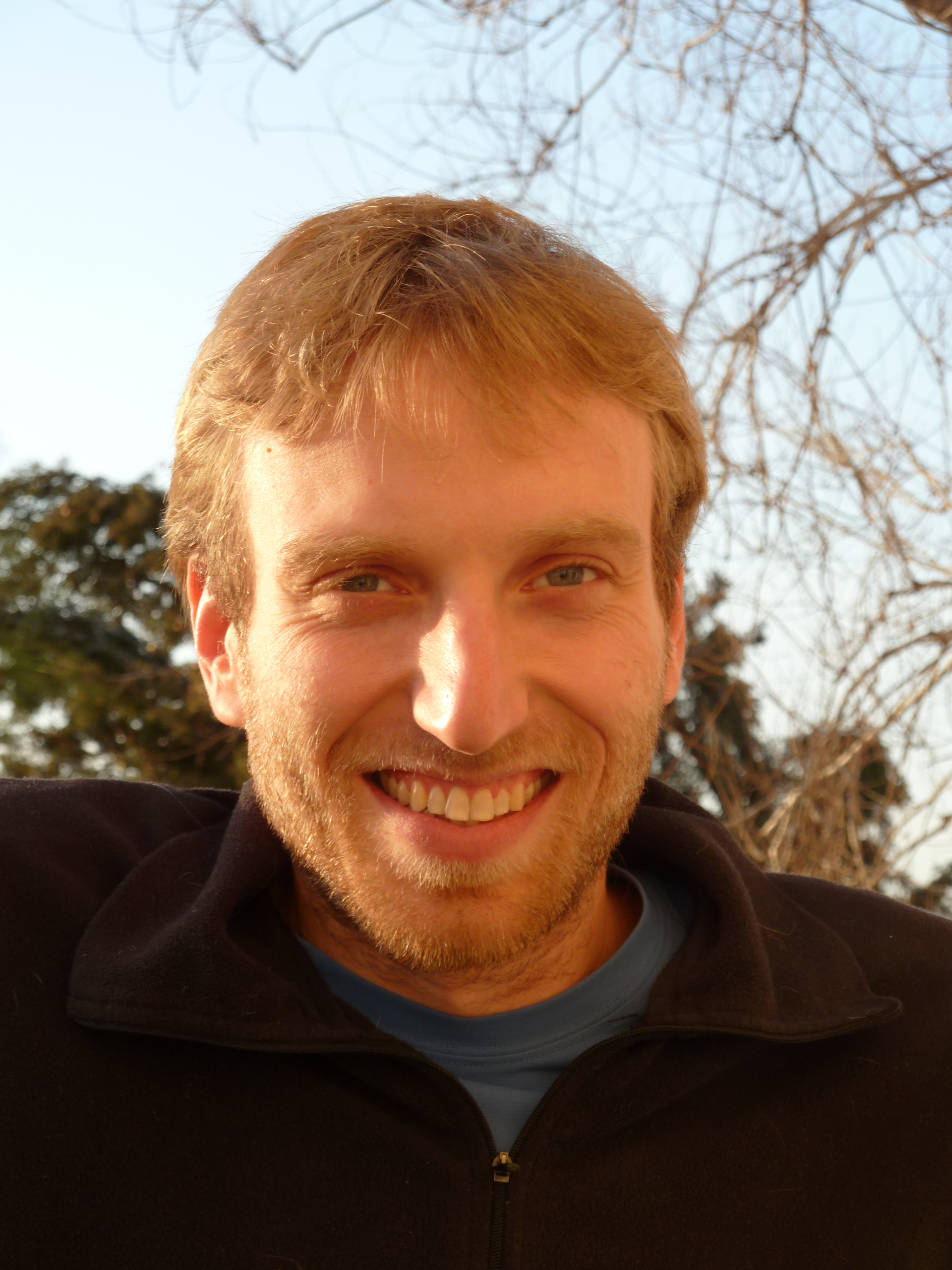}}]{Nir Shlezinger} (M’17-SM’23) is an assistant professor in the School of Electrical and Computer Engineering at Ben-Gurion University, Israel. He received his B.Sc., M.Sc., and Ph.D. degrees in 2011, 2013, and 2017, respectively, from Ben-Gurion University, Israel, all in electrical and computer engineering. From 2017 to 2019, he was a postdoctoral researcher at the Technion, and from 2019 to 2020, he was a postdoctoral researcher at the Weizmann Institute of Science, where he was awarded the FGS Prize for outstanding research achievements. His research interests include communications, information theory, signal processing, and machine learning.
\end{IEEEbiography}

\begin{IEEEbiography}[{\includegraphics[width=1in,height=1.25in,clip,keepaspectratio]{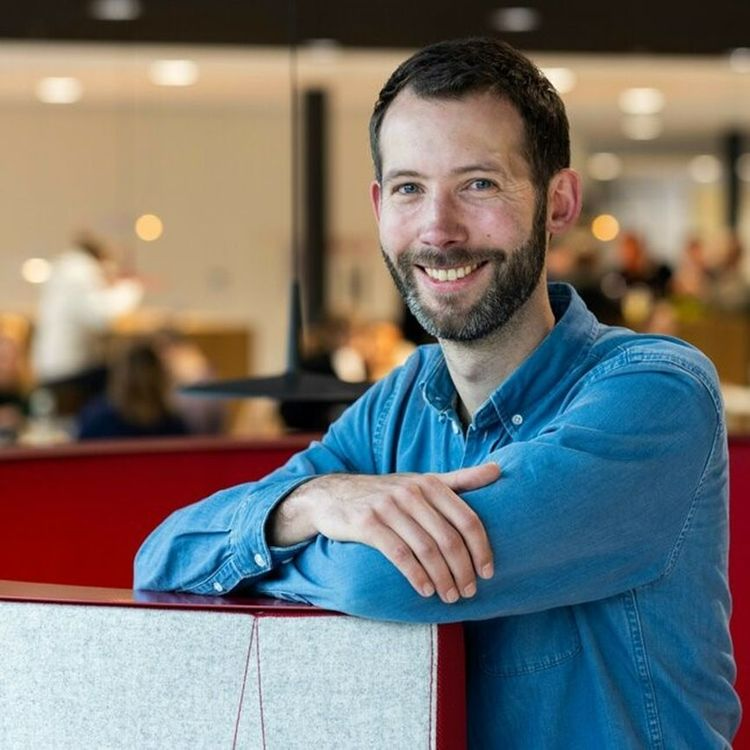}}] {Ruud van Sloun} is an Associate Professor at the Department of Electrical Engineering at Eindhoven University of Technology in the Netherlands. He received both his M.Sc. and Ph.D. degrees (cum laude) in Electrical Engineering from Eindhoven University of Technology in 2014 and 2018, respectively. From 2019 to 2020, he served as a Visiting Professor with the Department of Mathematics and Computer Science at the Weizmann Institute of Science in Rehovot, Israel. From 2020 to 2023, he was a Kickstart AI Fellow at Philips Research. He has been honored with an ERC Starting Grant, an NWO VIDI Grant, an NWO Rubicon Grant, and a Google Faculty Research Award. His current research interests include closed-loop image formation, deep learning for signal processing and imaging, active signal acquisition, model-based deep learning, compressed sensing, ultrasound imaging, and probabilistic signal and image reconstruction.
\end{IEEEbiography}

\begin{IEEEbiography}[{\includegraphics[width=1in,height=1.25in,clip,keepaspectratio]{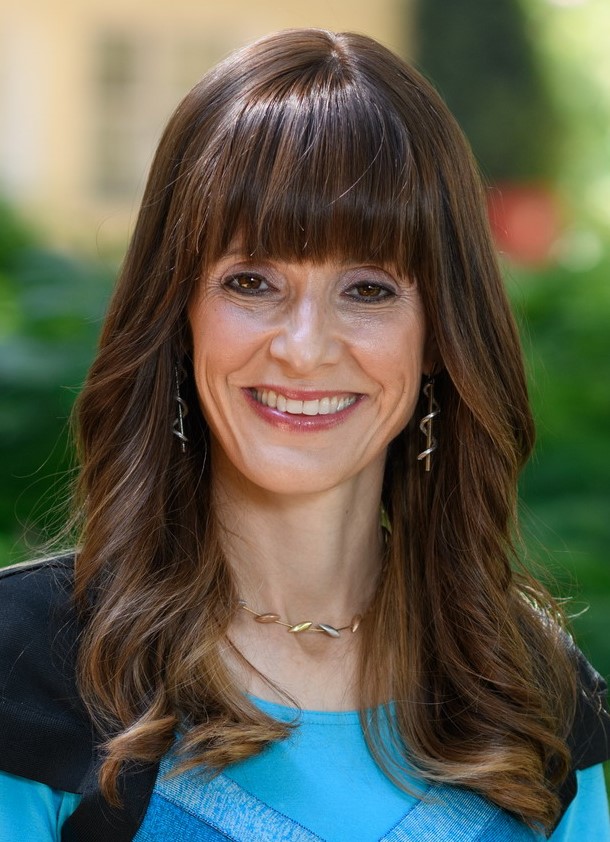}}]{Yonina C. Eldar}
	(S'98-M'02-SM'07-F'12)  received the B.Sc. degree in Physics in 1995 and the B.Sc. degree in Electrical Engineering in 1996 both from Tel-Aviv University (TAU), Tel-Aviv, Israel, and the Ph.D. degree in Electrical Engineering and Computer Science in 2002 from the Massachusetts Institute of Technology (MIT), Cambridge.
	
	She is currently a Professor in the Department of Mathematics and Computer Science, Weizmann Institute of Science, Rehovot, Israel. She was previously a Professor in the Department of Electrical Engineering at the Technion, where she held the Edwards Chair in Engineering. She is also a Visiting Professor at MIT, a Visiting Scientist at the Broad Institute, and an Adjunct Professor at Duke University and was a Visiting Professor at Stanford. She is a member of the Israel Academy of Sciences and Humanities (elected 2017), an IEEE Fellow and a EURASIP Fellow. Her research interests are in the broad areas of statistical signal processing, sampling theory and compressed sensing, learning and optimization methods, and their applications to biology, medical imaging and optics.
	
	Dr. Eldar has received many awards for excellence in research and teaching, including the  IEEE Signal Processing Society Technical Achievement Award (2013), the IEEE/AESS Fred Nathanson Memorial Radar Award (2014), and the IEEE Kiyo Tomiyasu Award (2016). She was a Horev Fellow of the Leaders in Science and Technology program at the Technion and an Alon Fellow. She received the Michael Bruno Memorial Award from the Rothschild Foundation, the Weizmann Prize for Exact Sciences, the Wolf Foundation Krill Prize for Excellence in Scientific Research, the Henry Taub Prize for Excellence in Research (twice), the Hershel Rich Innovation Award (three times), the Award for Women with Distinguished Contributions, the Andre and Bella Meyer Lectureship, the Career Development Chair at the Technion, the Muriel \& David Jacknow Award for Excellence in Teaching, and the Technion’s Award for Excellence in Teaching (two times).  She received several best paper awards and best demo awards together with her research students and colleagues including the SIAM outstanding Paper Prize, the UFFC Outstanding Paper Award, the Signal Processing Society Best Paper Award and the IET Circuits, Devices and Systems Premium Award, was selected as one of the 50 most influential women in Israel and in Asia, and is a highly cited researcher.
	
	She was a member of the Young Israel Academy of Science and Humanities and the Israel Committee for Higher Education. She is the Editor in Chief of Foundations and Trends in Signal Processing, a member of the IEEE Sensor Array and Multichannel Technical Committee and serves on several other IEEE committees. In the past, she was a Signal Processing Society Distinguished Lecturer, member of the IEEE Signal Processing Theory and Methods and Bio Imaging Signal Processing technical committees, and served as an associate editor for the IEEE Transactions On Signal Processing, the EURASIP Journal of Signal Processing, the SIAM Journal on Matrix Analysis and Applications, and the SIAM Journal on Imaging Sciences. She was Co-Chair and Technical Co-Chair of several international conferences and workshops. She is author of the book "Sampling Theory: Beyond Bandlimited Systems" and co-author of five other books published by Cambridge University Press.	
\end{IEEEbiography} 
\end{document}